# *Challenging the Machine*:
## Contestability in Government AI Systems
Recommendations and Summary of Workshop on Advanced Automated Systems, Contestability, and the Law

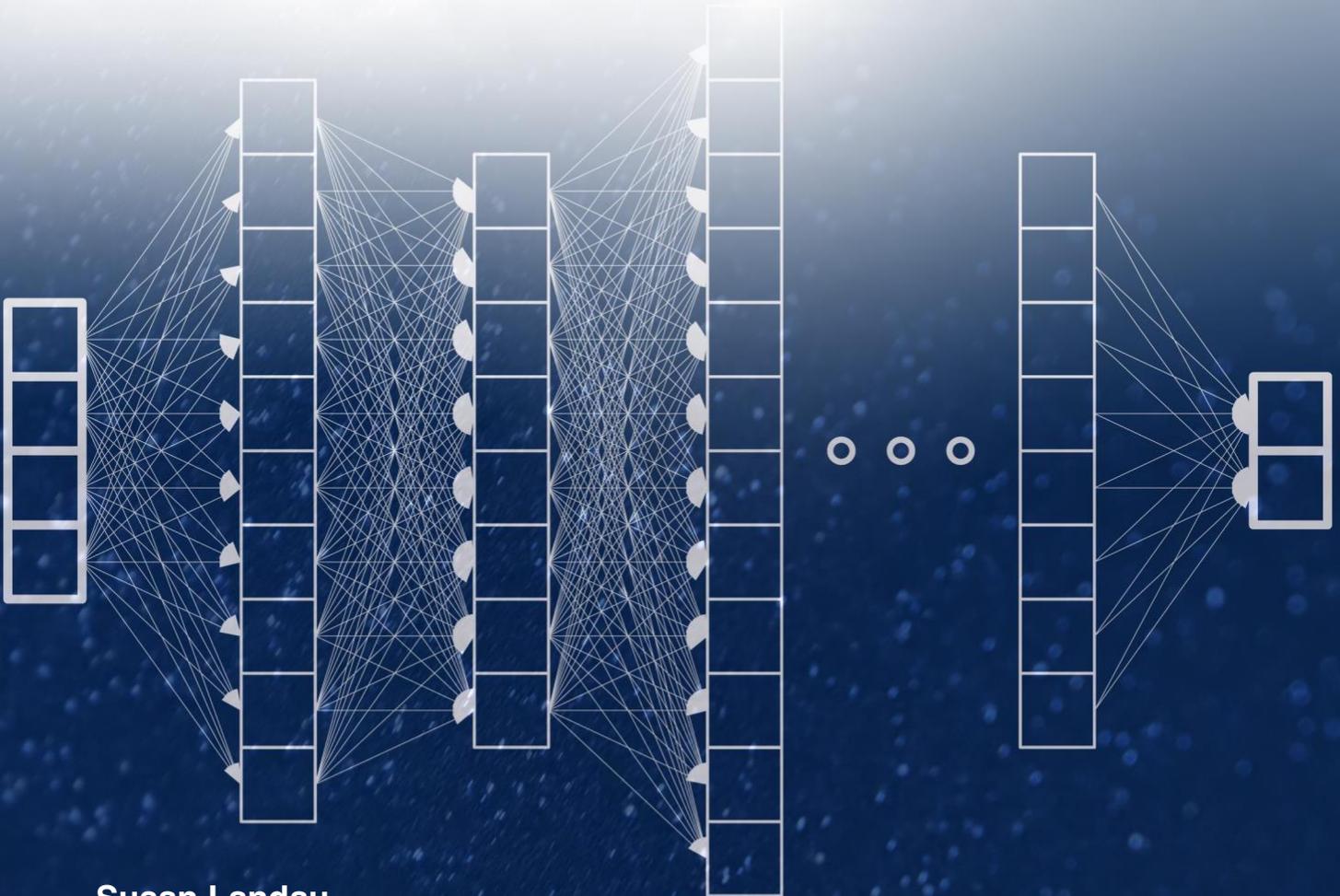

**Susan Landau**
**James X. Dempsey**
**Ece Kamar**
**Steven M. Bellovin**

**Workshop summary by Robert Pool, Susan Landau, and James X. Dempsey**
**June 2024**

# *Challenging the Machine*:

## Contestability in Government AI Systems

Recommendations and Summary of Workshop on
Advanced Automated Systems, Contestability, and the Law

**Susan Landau**
**James X. Dempsey**
**Ece Kamar**
**Steven M. Bellovin**

**Workshop summary by Robert Pool, Susan Landau, and James X. Dempsey**
**June 2024**

# Authors

**Steven M. Bellovin** is the Percy K. and Vida L. W. Hudson Professor of Computer Science at Columbia University, member of the Cybersecurity and Privacy Center of the university's Data Science Institute, and an affiliate faculty member at Columbia Law School. Bellovin does research on security and privacy and on related public policy issues. In his copious spare professional time, he does some work on the history of technology.

**James X. Dempsey** is senior policy advisor to the Stanford Program on Geopolitics, Technology, and Governance and a lecturer at UC Berkeley Law School. He is a former member of the U.S. Privacy and Civil Liberties Oversight Board. From 2015 to 2021, he was executive director of the Berkeley Center for Law & Technology. Prior to that, he was at the Center for Democracy & Technology, where he held a number of leadership positions, including executive director.

**Ece Kamar** is Vice President of Research and the Managing Director of the AI Frontiers Lab, where she leads research and development towards pushing the frontiers of AI capabilities. She is also Technical Advisor for Microsoft's Internal Committee on AI, Engineering and Ethics, leading efforts on reliability and safety of AI systems. Ece is an Affiliate Faculty in the Department of Computer Science and Engineering at the University of Washington. She is currently serving on the National Academies' Computer Science and Telecommunications Board.

**Susan Landau** is Bridge Professor of Cyber Security and Policy at the Fletcher School and the School of Engineering at Tufts University, where she does research at the intersection of privacy, surveillance, national security, and the law. She has testified before Congress on cybersecurity and encryption policy. She was previously senior staff privacy analyst at Google, distinguished engineer at Sun Microsystems, and held faculty positions at Worcester Polytechnic Institute, the University of Massachusetts, and Wesleyan University.

**Robert Pool** is an author and editor who specializes in writing about science and technology for a general audience. Earlier in his career he was a research news writer at *Science* magazine, and news editor at *Nature* before becoming freelance. The author of several books, Pool has been published in many of the world's leading science magazines, including not just *Science* and *Nature* but also *Discover*, *New Scientist*, and *Technology Review*. He has worked extensively with the National Academies of Science, Engineering, and Medicine.

# Table of Contents

# Part I: Recommendations for Government Development and Use of Advanced Automated Systems to Make Decisions about Individuals

**March 1, 2024**

**Susan Landau, James X. Dempsey, Ece Kamar, Steven M. Bellovin**

Contestability—the ability to effectively challenge a decision—is critical to the implementation of fairness. In the context of governmental decision making about individuals, contestability is often constitutionally required as an element of due process; specific procedures may be required by state or federal law relevant to a particular program. In addition, contestability can be a valuable way to discover systemic errors, contributing to ongoing assessments and system improvement.

The ability of a person to challenge a decision based in whole or part on an automated system can be limited due to insufficient information about the system, the technical opaqueness of the system, or broader problems with the right to contest as it exists on the books and on the ground (e.g., Citron, Wexler). As automated systems become more advanced due to the incorporation of machine learning and other artificial intelligence technologies, contestability may become even more difficult to achieve.

Yet it is not inevitable that advances in automated decision making will necessarily make contestability more difficult. To the contrary, as researchers and developers have shown, conscious choices made in system design can ensure that advanced automated systems enable meaningful contestability, perhaps even better than current systems do.

Here, we provide recommendations for the development and use of advanced automated systems to enable contestability.[1]

Contestability is not merely a best practice. Where decisions are being made about individuals, contestability is a requirement. Building on Executive Order 14110 on Safe, Secure and Trustworthy Development and Use of Artificial Intelligence (Oct. 30, 2023), the federal government should adopt binding standards on contestability applicable to the design, testing, implementation, and monitoring of advanced automated systems used by the federal government or in the administration of federally funded programs.

---

[1] When it comes to contestability, not all AI/ML techniques or capabilities are equal. Some may be incompatible with contestability; depending on the context, these should not be used.

Note that we use "system" here to encompass both technology and humans and both policy and data. Thus, when we say "system," we are encompassing such disparate systems as a hiring system, a system for allocating police resources, a system for granting or denying veterans' benefits or determining Medicaid eligibility, or a system for prioritizing enforcement actions. Note also that we use "program" here to refer to a governmental benefit, service, function or activity (e.g., a veterans' benefits program or an enforcement program), not to computer code or software.

In some systems, the automated technology will be public-facing and its impact on individuals will be immediately apparent, and the need for contestability will be clear. In other contexts, however, the automated technology will be operating in the background, supporting decisions that directly affect individuals. In such cases, careful consideration will need to be given to transparency to ensure that individuals can discern whether and how technology is affecting them, as a premise for contesting the results of such decision-support functions. In yet other contexts, automated analytics capabilities will improve backend operations of government agencies without implicating individual rights.

On January 24-25, 2024, with support from the National Science Foundation and the William and Flora Hewlett Foundation, we convened a diverse group of government officials, representatives of leading technology companies, technology and policy experts from academia and the non-profit sector, advocates, and stakeholders for a workshop on advanced automated decision making, contestability, and the law. Informed by the workshop's rich and wide-ranging discussion, we offer these recommendations. A full report summarizing the discussion is in preparation.

## Contestability by Design

Contestability is often discussed in terms of transparency, interpretability or explainability of the automated decision system, and contestability does require that meaningful explanations be provided to the user and the subject. However, contestability is more than just transparency, interpretability, or explanation. Moreover, the definitions of those terms are unsettled. In particular, "explanation" may have different meanings in AI/ML science versus in law or social science (see, e.g., Mittelstadt et al.).

To ensure contestability, those involved in the development of a system need to understand what can go wrong (what are the types and sources of errors) in a decision-making process.[2] Sometimes, a decision is wrong because the system was trained on inaccurate, biased, or irrelevant data. Other times, the automated elements of a system do not properly reflect the law, policy, or rules applicable to the particular government or private-sector program or activity. Sometimes the system is wrong in a specific decision because the data it uses about an individual is inaccurate or incomplete or accurate but irrelevant. Sometimes the system is wrong because a machine recommendation is unclear, and the human misinterprets the recommendation, making a wrong decision. Another possibility is that the tool is probabilistic and produces a certain percentage of errors even when everything is done perfectly.[3] Designing for the individual right to contestability should take into account all of these (and possibly other) error types. Nonetheless, although designing advanced automated systems to support a meaningful right of contestability is difficult, it is not impossible—and *it is often required by law.*

Contestability must be incorporated into a system by design, and considerations related to contestability must be addressed when initially determining whether to include automated decision making capabilities in a system and then throughout the development lifecycle, from conceptualization to implementation and must be examined after deployment in ongoing monitoring and improvement. At the same time, it is important to understand what level of contestability is sufficient for a given domain, and design accordingly. Note that no single practice or feature can ensure adequate contestability. Instead, contestability can be achieved only though application of a series of techniques, starting with design and impact assessment. Some of the techniques for ensuring contestability will relate to the technology, while others will relate to overall system policy.

## Recommendations: Building Contestability into Advanced Automated Systems Used in Government Decision Making about Individuals

***Recommendation 1: In contexts where contestability is required, government should ensure that adequate notice is provided that an automated decision-making system is being developed and being used.*** Notice is an essential prerequisite of contestability; this means both (i) notice to the public before the decision is made to adopt automated decision making for a system and then consultation as it is being developed and (ii)

---

[2] In referring to a system outputs as being "wrong" or "erroneous," we have in mind the language of the federal Administrative Procedure Act: "arbitrary, capricious, an abuse of discretion, or otherwise not in accordance with law; … in excess of statutory … authority, or limitations, or short of statutory right; … unsupported by substantial evidence …; or unwarranted by the facts … ." 5 U.S.C. § 706.

[3] Designers and users alike need to be aware of the error rate in such systems.

notice to individual subjects that their case has been decided based in whole or part on an automated process.

**Notice to the public must be adequate to allow for systemic challenges to a system**, which can identify problems before large numbers of individuals are unfairly affected.

**Notice to individuals must be understandable;** thus, notice must provide sufficient detail so that affected individuals and their representatives or advocates can understand how a decision was made and what a person must present to contest it.[4]

The degree of transparency necessary to support contestability will vary by context. Especially where the government fails to justify an outcome in a way that is understandable by the affected individual and his or her representative, it may be necessary for advocates and litigators to delve into how the system was constructed. Datasheets or model cards as documentation for how a system was built could enable contestability (Mitchell et al., Ehsan et al.), but in some cases a deeper examination of methods, criteria, code and data may be necessary along with expert analysis by those asserting challenges.

***Recommendation 2: Contestability must be incorporated into the system design, beginning with the decision whether to use an advanced automated system in a decision-making or a decision-supporting role.*** A risk assessment provides one opportunity early in the development lifecycle to surface issues of contestability.

Design choices—including decisions about which type of advanced automated system to deploy—can affect the ability of individuals or their representatives to effectively contest a decision based in whole or part on automated processes. If individuals can face adverse consequences (such as denial of a federal loan, loss or reduction of benefits, or increased scrutiny in a policing or enforcement context) from the decisions made or recommended by a system, then system designers must choose a technique or technology understandable (e.g., interpretable) by affected individuals and their representatives (if any).

***Recommendation 3: Designers should always consider the option of not deploying an advanced automated decision-making technique or technology, and they must build into each system affecting individuals the option of an off-ramp***

---

[4] In *Goldberg v. Kelly,* 397 U.S. 254, 267 (1970), the Supreme Court specified that "[t]he opportunity to be heard must be tailored to the capacities and circumstances of those who are to be heard." Thus the written notice that undergirds due process must be "in terms comprehensible to the claimant." *Ortiz v. Eichler*, 616 F. Supp. 1046, 1061 (D. Del. 1985), *aff'd*, 794 F.2d 889 (3d Cir. 1986). "An elementary and fundamental requirement of due process…is notice reasonably calculated, under all the circumstances, to apprise interested parties of the pendency of the action and afford them an opportunity to present their objections." *Mullane v. Cent. Hanover Bank & Tr. Co.*, 339 U.S. 306, 314 (1950).

(that is, the ability, even though it is a very hard decision to make if considerable resources have been expended in development, to exit from the use of automated decision making) if a system is determined, after deployment, to be not sufficiently contestable. Similarly, agencies should carefully monitor the volume and nature of challenges to decisions made or aided by deployed systems; a high rate of challenges may suggest flaws in the system, requiring redesign or even decommissioning.

***Recommendation 4: Design consultations and testing should include different types of system participants, including operators, end users, decision makers, and decision subjects.*** [5] Managers familiar with the legal and policy aspects of a governmental program or function must work together with experts in AI or other relevant advanced analytic technologies at all stages of system lifecycle from initial conceptualization through post-deployment assessment and revision.

To ensure contestability, there must be participation throughout a project's life cycle not only from experts in machine learning but also data scientists, statisticians, and experts in such fields as human-computer interface (HCI), design, sociology, cognitive psychology, linguistics, law, and public policy. Other areas of expertise, including criminology, health care, and economics, may also be needed depending on the type of system being developed.

***Recommendation 5: Stakeholders who will be directly affected by an advanced automated system must be involved or represented at all stages of its development and use.*** This includes from the initial discussions of the scope of the system, through design, development, iterative testing phases, deployment, post-deployment assessment, and revisions/updates of the system.[6] Special effort must be made to include persons who are disabled, lack English language skills, or have otherwise traditionally been disadvantaged from full participation in government processes.[7] Because they could offer unique systemic perspectives, advocates such as lawyers handling challenges in the program or function at issue should also be included.

***Recommendation 6: The contestability features of a system must be stress tested with real world examples and scenarios before field deployment.*** [8] This also holds

---

[5] See NIST RMF Playbook, Measure 2.8.

[6] In the case of systems affecting children and perhaps some other categories of persons, participation can be through representatives or advocates.

[7] Though there is expertise is available in how to conduct inclusive project design consultations, it is critical that such efforts be done in a way that actually engages individuals who traditionally were not involved in such consultations (see, e.g., Sloane, 2022).

[8] Individuals in the test set should not end up worse off than if their case had been handled by the old system.

true for non-trivial modifications of a deployed system. Before field deployment, it is critical to conduct pilots on real data and in consultation with individuals in all the different types of actual communities in which the system will be deployed. Among other issues, testing should consider whether there are misunderstandings between technology developers on the one hand and lay users and data subjects on the other hand. After deployment, systems may be updated and the scenarios in which they are used may change, so there must be ongoing evaluation of the system's operation.

***Recommendation 7: Contestability processes should be equally accessible and usable by people with different backgrounds,*** including—but not limited to—different cultures, languages, education levels, and incomes. Having robust contestability processes is not sufficient in itself; those processes must be widely accessible and usable.

Contestability of any system depends in large part on who will be contesting decisions, what resources they have available to them, and what barriers they face. Governments already struggle to provide clear explanations of what people need to prove to get or keep benefits. Low-income people especially find it difficult to contest government decisions such as benefit cuts or denials. Many affected persons do not have the money needed to hire a lawyer, and the capacity of organizations providing free legal services is severely limited. Even without adoption of advanced technology, formal processes to contest can be confusing, involve burdensome paperwork requirements, and require time away from work or caregiving obligations. As automated systems are designed and implemented, government entities should simplify, streamline and explain more clearly their application and appeal processes.

***Recommendation 8: Reproducibility is crucial.*** In order to be contestable, a system must be able to reproduce a given decision (that is, the same inputs must produce the same outputs at the time of challenge). Thus, there needs to be version control and thorough recordkeeping of the systems being used (Aler Tubella et al.), including of the parameters of the models created from training data.

Governments should fully document the relevant data (including training data), statistical modelling, impact projections, system changes, and assumptions involved in the design, development, and testing of any system prior to deployment. This information will aid the agency in making decisions about how the system will be deployed, if at all. In addition, it will aid post-deployment accountability efforts by allowing for internal and public review, strengthening appeal processes, and facilitating agency or legislative oversight efforts.

***Recommendation 9: The automated features of a system should never be allowed to supplant or displace the criteria specified in law for any given program or function.*** The convenience of the programmers or even the technical possibilities of automated systems must take a back seat to what the law requires. For example, if the legal standard is "medical necessity," the factors or criteria considered by the automated process should not be presumed to be the only way to demonstrate medical necessity. If an individual has a due process right to contest a decision, that must include the right to present to the reviewing

authority factors or criteria relevant to the legal standard that were not included in the automated process.

***Recommendation 10: Additional research could help government agencies design and implement meaningful contestability processes.*** Such work might include a clearer understanding of understandability, better knowledge of the social impact of advanced automated decision making on various communities (including businesses and organizations), whether new legal theory is needed to ensure contestability of automated systems, and evaluation of the risks of "out-of-control" contestability (e.g., when AI bots create a deluge of comments to government agencies in response to a rulemaking or other public consultation).

Although new research could improve the design and implementation of contestability in advanced automated decision-making systems, this does not obviate any of our other recommendations; the actions we recommend must proceed even as contestability capabilities improve due to new understanding and research.

## Contestability Must be Addressed in the Procurement Process

The procurement process—the nuts and bolts of government contracting—is critical because many automated decision-making systems will be designed and built (and may be managed as a service) for the government by contractors. Thus, the procurement process must be part of the government's efforts in ensuring contestability.

***Recommendation 11: The procurement process should be leveraged to ensure that advanced automated systems genuinely enable contestability.*** Solicitations and contracts must clearly require contractors to deliver contestability as a core system feature. Contractors should not be allowed to use assertions of trade secrecy or other intellectual property claims to frustrate contestation.[9] As OMB develops procurement guidelines for AI pursuant to Section 10.1(d)(ii) of the AI Executive Order, it should include contestability as a core element in procurement processes; a set of unified guidelines that all procurement officers must adhere to across agencies would be powerful and actionable.

***Recommendation 12: Federal officials should ensure that contestability is required of the states implementing federal programs and of private companies whose systems, such as credit scoring, are used by the government in contexts***

---

[9] The logic of an advanced automated process may or may not be useful for contestability, but that question should not be pretermitted by IP claims.

***affecting individuals.*** Federal officials can do this by using their approval and oversight authority over programs that are administered partly by states or receive federal funds.

***Recommendation 13: The government should develop expertise to ensure rights-impacting advanced automated systems are contestable.*** The type of socio-technical expertise needed by agencies designing, procuring, and using advanced automated systems is hard to come by. Training within individual agencies, as called for under EO 14110, may be unnecessarily narrow—and thus ultimately ineffective—given the rapid ongoing evolution of the technology, the constitutional and thus cross-disciplinary foundations of contestability, and the generalizable nature of processes for stakeholder consultation as well as for government procurement. It is necessary to ensure that the government workforce has sufficient understanding regarding the development and use of these socio-technical systems.

***Recommendation 14: The federal government should develop a centralized training function in development and assessment of advanced automated system for government programs.***[10] Among others in government who will need to understand the implications of advanced automated systems, agency staff who will be adjudicating appeals, including administrative law judges, should be specially trained to identify the risks of automated systems.

***Recommendation 15: There should be formal—and informal—ways set up to ensure sharing knowledge gained in the development, procurement, and use of these systems.*** Many of the efforts within federal, state, and local governments will have common characteristics in terms of the communities they are serving and the challenges they face. The Chief AI Officers of the agencies who will be appointed pursuant to Section 10 of EO 14110 should take this on as one of their responsibilities, and the interagency council established under the same section of the EO should serve as one vehicle for such knowledge sharing.

***Recommendation 16: At all levels of an agency, from the agency head to the procurement officer to the case worker or other person interfacing with affected***

---

[10] This recommendation goes beyond the provisions on AI talent in EO 14110 on Safe, Secure and Trustworthy Development and Use of Artificial Intelligence (Oct. 30, 2023). Sections 10.2 (a)-(f) of the EO address measures the government will take in recruitment and hiring, but the talent pool will always be too small, as universities and other educational institutions are unlikely by themselves to produce enough graduates with the skills needed by the private sector and government. Likewise, Section 5.1 of the EO seek to improve the nation's ability to attract AI talent from abroad, but there too the U.S. is in competition with government of other nations and the private sector of other countries also facing the same talent shortage. Section 10.2(g) of the EO recognizes that the government will have to undertake training itself, but it leaves that training to each agency head. Just as the federal government has established centralized training facilities for other skills, such as languages, cryptography, and law enforcement, it should establish a centralized AI governance institute.

***individuals, officials need to understand, at a level appropriate to their roles, the benefits, limitations, and risks of automated decision making.*** Likewise, at all levels, officials need to understand, at a level appropriate to their roles, the capabilities of automated decision making to deliver on stated goals. That presumes that officials are clear on their goals for the system being developed (for example, whether the intent is to enroll more people in a program or fewer).

***Recommendation 17: In order to build sufficient expertise within agencies, as an early effort, agencies should attempt some low-risk, high-gain systems.*** A low-risk system is one that does not have adverse effects on individuals if there are malfunctions.

## Notes on the Scope of the Recommendations

The foregoing recommendations were informed by discussions during the final session of the workshop. That session also raised several points about the scope and applicability of any principles that might be adopted to guide the development, procurement, and use of advanced automated decision-making technologies.

The workshop's intent was to consider how to ensure contestability of systems using the technologies that would be deployed five to ten years in the future. Thus, much of the discussion focused on AI/ML systems. Yet many of the systems that will be used ten years from now and even later will not be AI/ML based, but rather will be based on simpler automated techniques. For this reason, the recommendations are deliberately agnostic as to the type of technology employed in advanced automated decision-making systems.

Another concern was exactly to which type of systems the recommendations should apply. Should such requirements for contestability apply just to public-facing systems or should they also apply to back-office systems, such as those related to an agency's financial management? There was general agreement that, given our focus on contestability, the distinction should be not between public-facing and back-office systems, but between those systems that affect individuals and those that do not; the recommendations should apply only to the former.

A related issue was whether the systems of interest are limited to those where the algorithm or model makes a final decision about an individual or whether concerns about contestability also apply to systems that support human decision making. Some might say that there is a fundamental difference between a system in which the technology makes a decision about eligibility for a program versus a system where the technology processes the data and produces some output, which is then used in the decision, but the ultimate decision is left up to a human. It was noted, however, that in such cases decision-support technology may effectively function as decision-making technology. Suppose, for example, that an automated system assigns a credit score of 580 to a person. If the programmatic rule applied by a loan officer examining the application says that loans can only be approved for persons with scores above 580, then it is not a real distinction to say that a human makes the final decision, "FICO score of 580: loan denied." This is effectively a decision made by the algorithm. Thus, the recommendations should apply not only to decision-making technologies, but also to decision-support technologies.

These points should inform decision makers at all levels as they implement and use advanced automated systems for decision making about individuals.

# Resources

# Part II: A Short History of Due Process and Computer Decision Making About Individuals

## James X. Dempsey and Susan Landau

Machine learning, especially deep learning, with its seemingly inexplicable results, poses issues of great concern, especially when the results directly affect human lives. As developers, policymakers, and society at large consider how to tame these machines, it is important to recall that this is not the first time that "The computer said so" has had the potential to directly harm individuals.

In the 1960s, the IBM 360 deeply shifted the balance of power between individuals and institutions in both the private sector and the government. As one of us wrote earlier, "Because the machine could handle large amounts of data, it was widely adopted by banks and other financial institutions . . . [U]sing computers and information sharing across financial institutions, banks could carry out their own calculations to assess the risk of default for a potential borrower. Suddenly banks had the upper hand, using increased access to data to decide whom to back and whom to charge higher interest rates" (Landau).

Policymakers responded with laws to give individuals certain rights regarding computerized decision making. "[O]ne result was the 1972 Fair Credit Reporting Act, which gave consumers access to and rights over their credit reports. This effort to redress the balance between people and lending institutions was the first of a stream of laws and regulations that would seek to protect people as computer technology changed their lives" (Landau).

The adoption of automated data processing also led to the highly influential study sponsored by the federal Department of Health, Education and Welfare, *Records, Computers, and the Rights of Citizens* (1973). This, in turn, led to the adoption of the federal Privacy Act, giving individuals the right to access their own records. At the same time, the Supreme Court elevated the principle of due process across a wide range of government decision-making contexts. In a series of cases, the Court held that parties whose rights are to be affected by government decisions are entitled to be heard and in order that they may enjoy that right they must first be notified. Moreover, the right to notice and an opportunity to be heard must be granted at a meaningful time and in a meaningful manner. *Mathews v. Eldridge*, 424 U.S. 319 (1976); *Fuentes v. Shevin*, 407 U.S. 67 (1972); *Goldberg v. Kelly,* 397 U.S. 254 (1970).

These due process rights were never definitively articulated. The Supreme Court said that what information should be provided to the individual and what the right to be heard meant in practice would vary from situation to situation, depending on the words of any applicable statute plus a balancing test assessing the private interest at stake, the risk of an erroneous decision, and the probable value, if any, of additional or substitute procedural safeguards, as compared to the government's interest, including the fiscal and administrative burdens that the additional or substitute procedural requirement would entail. Contestability, even when constitutionally mandated, remained a challenge for individuals with limited resources and difficult lives.

Artificial intelligence (AI) and machine learning's (ML) remarkable advances have created a new imbalance. As Danielle Citron observed, "The twenty-first century's automated decision-making systems bring radical change to the administrative state that last century's procedural structures cannot manage" (Citron).

Some of the negative implications of AI/ML have been well documented (although not resolved). Reliance on training data for ML systems, including historical data, has been shown in many cases to introduce highly biased results (see, e.g., Angwin et al.; Benjamin; Buolamwini and Gebru; Chouldechova et al.). While the underlying principle that individuals should have the right to challenge decisions based on automated processing is not under question, the actual ability to query the results of such processing and understand how the results are arrived could be rendered non-existent unless contestability is built in by design. For example, the large number of variables used in the computation can preclude any type of transparency in how the results are derived (see, e.g., Hirsch; Tulio). Depending on how a system is designed, a right that exists under law may not actually be one that exists in practice.

As AI systems proliferate, there have been two approaches to the policy issues: application of existing laws (including the federal Administrative Procedure Act and constitutional principles) and development of new ones. Multiple federal and state regulatory agencies have issued guidance stating that existing anti-discrimination laws apply to AI-enabled decisions. For example, in April 2023, the Consumer Financial Protection Bureau, Department of Justice, Equal Employment Opportunity Commission, and Federal Trade Commission issued a joint statement reminding all "existing legal authorities apply to the use of automated systems and innovative new technologies." Meanwhile, welfare rights advocates have successfully used pre-AI language in federal benefits laws to challenge specific uses of AI-based systems (see, e.g., *Barry v. Lyon*, 834 F.3d 706 (6th Cir. 2016); Richardson et al.; Brown et al.).

The other approach is to develop new laws specifically aimed at automated decision making. Here, much of the action has been at the state level.[11] As of March 13, 2024, twelve states have laws addressing certain uses of automated decision making under the concept of "profiling." These state laws generally require businesses to allow consumers to opt-out of profiling in furtherance of decisions that produce legal or similarly significant effects concerning the consumer. Many also require data controllers to conduct a data protection assessment of certain risky uses of profiling. A few states give consumers the right to access to information about the technology.[12] But none of them gives consumers the right to challenge the accuracy of decisions made by automated means. Also, many of the state laws have broad exceptions or exclusions; most for example, do not apply to financial data, or credit data, or in the employment context. Also, the laws focus on consumer privacy, so they regulate businesses, not government agencies.

---

[11] Marijn Storm and Marian A. Waldmann Agarwal, AI Trends for 2024 - U.S. State Consumer Privacy Laws' Focus on Automated Decision-Making and Profiling, MoFo (Dec. 22, 2023); Rachel Wright, Artificial Intelligence in the States: Emerging Legislation, Council of State Governments (Dec. 6, 2023).

A few states have begun to address governmental uses of AI. For example, a California statute, AB 302, requires the Department of Technology to conduct a comprehensive inventory of all high-risk automated decision systems being used or proposed for use by any state agency. The inventory must describe the decisions the automated decision system can make or support and the measures in place, if any, to mitigate the risks, including the risk of inaccurate, unfairly discriminatory, or biased decisions. One municipality, New York City, has adopted a local law requiring employers to conduct a bias audit of any automated employment decision tool and provide notice to applicants and employees explaining how the tool influences employment decisions.[13] So far, though, contestability seems not to be a focus of state or local laws.

We convened our January 2024 workshop within this context. Building on the history of efforts to guarantee fundamental rights in the face of technological change, and informed by the unique challenges posed by artificial intelligence, the workshop focused on identifying specific ways to ensure that government advanced automated systems involving decision making affecting individuals were designed to enable effective contestability.

---

[12] For example, regulations in Colorado give consumers the right to a non-technical, plain language explanation of the logic used in the profiling process and the right to correct or delete the personal data used. Colorado Privacy Act Rules, 4 CCR 904-3, Rule 9.04.

[13] NYC Department of Consumer and Worker Protection, Automated Employment Decision Tools: Frequently Asked Questions (2023).

# Resources

Richardson, Rashida, Schultz, Jason M., and Southerland, Vincent M., "Litigating Algorithms: 2019 US Report," AI Now Institute (2019).

Rudin, Cynthia, Stop explaining black box machine learning models for high stakes decisions and use interpretable models instead, *Nature Machine Intelligence* (2019).

Tulio Ribeiro, Marco, Singh, Sameer, and Guestrin. Carlos, " 'Why should I trust you?' Explaining the predictions of any classifier," *Proceedings of the 22nd ACM SIGKDD International Conference on Knowledge Discovery and Data Mining*, pp. 1135–1144 (2016).

Records, Computers and the Rights of Citizens, Report of the Secretary's Advisory Committee on Automated Personal Data Systems, U.S. Department of Health, Education & Welfare (1973).

Wexler, Rebecca, Life, Liberty, and Trade Secrets: Intellectual Property in the Criminal Justice System, 70 *Stanford Law Review* 1343 (2018).

White House, [Executive Order 14110](), on the Safe, Secure, and Trustworthy Development and Use of Artificial Intelligence, 88 Fed. Reg. 75191 (October 30, 2023).
16

# Part III: Workshop Summary

**Robert Pool, Susan Landau, and James X. Dempsey**

## Chapter 1: Introduction

In an October 2023 executive order (EO), President Biden issued a detailed but largely aspirational road map for the safe and responsible development and use of artificial intelligence (AI).[14] The challenge for the January 24-25, 2024 workshop was to transform those aspirations regarding one specific but crucial issue—the ability of individuals to challenge government decisions made about themselves—into actionable guidance enabling agencies to develop, procure, and use genuinely contestable advanced automated decision-making systems. While the Administration has taken important steps since the October 2023 EO, the insights garnered from our workshop remain highly relevant, as the requirements for contestability of advanced decision-making systems are not yet fully defined or implemented.

The workshop brought together technologists, members of government agencies and civil society organizations, litigators, and researchers in an intensive two-day meeting that examined the challenges that users, developers, and agencies faced in enabling contestability in light of advanced automated decision-making systems. To ensure a free and open flow of discussion, the meeting was held under a modified version of the Chatham House rule.[15] Participants were free to use any information or details that they learned, but they may not attribute any remarks made at the meeting by the identity or the affiliation of the speaker. Thus, the workshop summary that follows anonymizes speakers and their affiliation. Where an identification of an agency, company, or organization is made, it is done from a public, identified resource and does not necessarily reflect statements made by participants at the workshop.

One exception was made to the Chatham House rule. Michael Hawes of the U.S. Census Bureau described his agency's experience in presenting its use of differential privacy to the public and various stakeholders. Because the speaker obviously had detailed knowledge about discussions at the Census Bureau, it made no sense to keep his identity secret. But, with the exception of Hawes's presentation, none of the statements, ideas, arguments, or suggestions reproduced in this summary are attributed to individual speakers.

---

[14] White House, Executive Order 14110, on the Safe, Secure, and Trustworthy Development and Use of Artificial Intelligence, 88 Fed. Reg. 75191 (October 30, 2023) https://www.federalregister.gov/documents/2023/11/01/2023-24283/safe-secure-and-trustworthy-development-and-use-of-artificial-intelligence

[15] Chatham House rule: "When a meeting, or part thereof, is held under the Chatham House rule, participants are free to use the information received, but neither the identity nor the affiliation of the speaker(s), nor that of any other participant, may be revealed." https://www.chathamhouse.org/about-us/chatham-house-rule (accessed April 1, 2024).

We note that, while the workshop largely focused on AI/ML systems, the challenges of present-day automated systems formed a significant aspect of our discussions, and the recommendations presented in Part I apply to *all* advanced automated decision-making systems.

The workshop was intended to develop an understanding of when contestability is needed in advanced automated decision-making systems and to elucidate ways to enable contestability. But knowledge was generated by people in the room; as such, this workshop summary should *not* be taken as a comprehensive review of all current work on contestability and advanced automated decision-making system. This is especially the case given the rate at which the literature in the field is growing. In tandem, we note that we had more limited industry participation than we had sought. The lack of participation was not due to lack of invitations to industry players. Despite multiple follow-ups, and despite the importance of the topic and our making clear our desire for industry input, key developers of AI systems did not show up.

We began the report of this workshop with our Recommendations, which is actually the outcome of our work. Thus, this workshop summary is organized as follows. Chapter 2 offers some basic explanations of contestability as it applies to the decisions of federal agencies, what it means to "understand" an AI or ML system, and how contestability works—and fails to work—when advanced automated systems play a role in decision making about individuals. Chapter 3 discusses agency decision making from the perspective of those individuals affected by the decisions. Chapter 4 provides an overview of the current uses of advanced automated systems in federal and state governmental agencies. Chapter 5 focuses on industry responses to risks posed by AI and ML tools. Chapter 6 is not about advanced automated systems per se; instead, it provides an object lesson of how an agency can effectively communicate about and explain a new and poorly understood technology with the individuals and other stakeholders who are affected by that technology. Chapter 7 examines the importance of procurement in incorporating a new technology in an agency and, most important, in making sure that the new technology fits the needs of that agency and will provide what it is supposed to provide. Rather than ending with conclusions—those are authored by the four co-organizers of the workshop and are presented in Part I—Chapter 8 of the workshop summary offers several underlying reflections from the meeting.

The workshop was organized by Steven M. Bellovin, James X. Dempsey, Ece Kamar, and Susan Landau, with support from the National Science Foundation and the William and Flora Hewlett Foundation. It was held January 24-25, 2024, at the National Science Foundation, in Arlington VA. Participants are listed in the Appendix.

# Chapter 2: Contestability and Advanced Automated Systems

Ensuring the contestability of advanced automated systems requires attention to two very different areas: (i) the constitutional, statutory, and regulatory principles that mandate contestability of certain government decisions and (ii) the technology of AI and ML. This chapter offers three tutorials to prepare the reader for the rest of the report. It begins with a short constitutional and administrative law primer on contestability. Next, it examines what it means to "understand" the decisions made by an ML system. Then the chapter addresses contestability in the context of the technology underlying advanced automated systems, and how understanding the decisions made by an ML system is crucial to the ability to contest a decision. The chapter concludes with a brief discussion of future concerns.

We remind the reader that the discussion here represents viewpoints expressed at the workshop rather than a comprehensive review of all potential approaches to contestability in advanced automated decision-making systems.

***THE CONSTITUTION AND ADMINISTATIVE LAW: FOUNDATIONS OF CONTESTABILITY***

What is contestability, and why are certain actions and decisions of federal agencies and other government actors required to be contestable? The answers can be found in the U.S. Constitution and various statutes and regulations.

## *Constitutional Due Process*

The Fifth Amendment to the U.S. Constitution prevents the federal government from depriving any person of "life, liberty, or property, without due process of law." The Fourteenth Amendment applies the same language to the states. The due process right applies in the criminal justice context and in many regulatory contexts where an individual or a corporation may face fines or other penalties, but the courts have also held that many governmental programs such as veterans benefits or disability programs create a property right, to which due process attaches.[16]

The two foundational elements of due process are notice and an opportunity to be heard. Beyond that, however, the contours of due process are not fixed. "[D]ue process is flexible, and calls for such procedural protections as the particular situation demands."[17] In determining whether sufficient due process has been provided in a particular situation, the Supreme Court has adopted a balancing test that considers three factors: "the private interest that will be affected by the official action; second, the risk of an erroneous deprivation of such interest through the procedures used, and the probable value, if any, of additional or substitute procedural safeguards; and, finally, the Government's interest, including the function involved and the fiscal and administrative burdens that the additional or substitute procedural requirement would entail."[18]

---

[16] *Goldberg v. Kelly*, 397 U.S. 254 (1970).

[17] Morrissey v. Brewer, 408 U. S. 471, 481 (1972).

Especially pertinent, given the complexity of many AI/ML systems and the inscrutability of some, is the Supreme Court's insistence on understandable notice: "An elementary and fundamental requirement of due process … is notice reasonably calculated, under all the circumstances, to apprise interested parties of the pendency of the action and afford them an opportunity to present their objections."[19]

### *Legislative Requirements*

In addition, there are legislative requirements to provide notice and an opportunity to be heard. The first source of such rights are the individual organic statutes creating agencies and defining benefits programs. For example, 38 U.S.C. §301 created the Department of Veterans Affairs, 38 U.S.C. §1710 defines the broad standards of eligibility for hospital, nursing home, and domiciliary care, and a number of statutory provisions spell out the due process rights of veterans, including higher-level review by the agency of original jurisdiction (38 U.S.C. § 5104B), options following decision by the agency of original jurisdiction (38 U.S.C. § 5104C), and the jurisdiction and procedures of the Board of Veterans Appeals options following decision by agency of original jurisdiction (38 U.S.C. Chapter 71).[20]

Another key source of due process rights is the federal Administrative Procedure Act (APA), which is an administrative "super statute" in that it sets procedural requirements and standards for certain kinds of actions across all federal agencies. Under the APA, there are two main ways in which agencies act—through rulemaking or adjudication—and each of these can be either formal or informal.

A rule is defined as "an agency statement of general or particular applicability and future effect designed to implement, interpret, or prescribe law or policy." In federal agencies today there is almost never formal rulemaking; instead, informal rulemaking, which is also known as "notice and comment rulemaking," is the norm. The APA specifies the procedure for informal rulemaking, in which an agency must publish a notice of proposed rulemaking, offer interested individuals an opportunity to comment, and respond to substantive comments, before adopting and publishing the final rule.

---

[18] Mathews v. Eldridge, 424 U.S. 319, 334-35 (1976).

[19] Mullane v. Central Hanover Bank & Trust Co., 339 U.S. 306, 314 (1950).

[20] For other examples of the statutory sources of due process rights, see David Freeman Engstrom, Daniel E. Ho, Catherine M. Sharkey & Mariano-Florentino Cuéllar, *Government by Algorithm: Artificial Intelligence in Federal Administrative Agencies* (Feb. 2020) (report to the Admin. Conf. of the U.S.), https://www.law.stanford.edu/wp-content/uploads/2020/02/ACUS-AI-Report.pdf, at pp. 37-39 (adjudications in programs under the Social Security Administration); pp. 46-47, 50 (decisions by the Patent Office).

However, there are also many very important agency actions, such as setting priorities or issuing guidance, that fall outside the purview of the APA. An interesting question as agencies adopt AI and ML for increasingly many functions (such as using a chatbot to answer questions about student financial aid or prioritizing the allocation of enforcement resources) is whether efforts to ensure contestability should focus only on those agency functions that fall under the purview of the APA or whether contestability should also apply to agency decisions to adopt AI/ML technologies for functions that fall outside the scope of the APA. For example, a decision to not enforce a regulation is presumptively unreviewable because enforcement choices are at the discretion of the agency. What happens if AI/ML technologies play a role in these unreviewable decisions? [21]

---

[21] The March 2024 memorandum issued by the Office of Management and Budget (OMB) begins to answer this question. Memorandum M-24-10, Advancing Governance, Innovation, and Risk Management for Agency Use of Artificial Intelligence (March 28, 2024), https://www.whitehouse.gov/wp-content/uploads/2024/03/M-24-10-Advancing-Governance-Innovation-and-Risk-Management-for-Agency-Use-of-Artificial-Intelligence.pdf [permalink: https://perma.cc/7GPP-9EAY]. It requires each agency (except the Department of Defense and the intelligence agencies) to individually inventory each of its AI use cases at least annually, submit the inventory to OMB, and post a public version on the agency's website. Further, the memo requires:

> Agencies must ensure, to the extent consistent with applicable law and governmentwide guidance, including concerning protection of privacy and of sensitive law enforcement, national security, and other protected information, that the AI's entry in the use case inventory provides accessible documentation in plain language of the system's functionality to serve as public notice of the AI to its users and the general public. Where people interact with a service relying on the AI and are likely to be impacted by the AI, agencies must also provide reasonable and timely notice about the use of the AI and a means to directly access any public documentation about it in the use case inventory. [Footnote omitted.]

Moreover, the memo requires agencies to provide an opportunity for public comment:

> Consistent with applicable law and governmentwide guidance, agencies must consult affected communities, including underserved communities, and they must solicit public feedback, where appropriate, in the design, development, and use of the AI and use such feedback to inform agency decision-making regarding the AI. The consultation and feedback process must include seeking input on the agency's approach to implementing the minimum risk management practices established in Section 5(c) of this memorandum, such as applicable opt-out procedures. … Agencies are strongly encouraged to solicit feedback on an ongoing basis from affected communities in particular as well as from the public broadly, especially after significant modifications to the AI or the conditions or context in which it is used. In the course of assessing such feedback, if an agency determines that the use of AI in a given context would cause more harm than good, the agency should not use the AI. [Footnote omitted.]

This is not full APA rulemaking, and affected persons cannot enforce the procedures established under an OMB memo, but it does require a form of notice and comment for agency actions that fall outside the APA.

Adjudication, the second way in which agencies act under the APA, is defined broadly as any "agency process for the formulation of an order."[22] Generally speaking, an adjudication is a decision affecting a specific person or entity. If the statute creating a government function says that an adjudication must be "determined on the record after opportunity for an agency hearing," that is called a "formal" adjudication, and the APA defines what such a proceeding must consist of. Formal adjudications look like trials, with opportunities for oral presentation and to confront witnesses. Other types of adjudications, called "informal," are not addressed specifically in the APA; the sole source of procedural due process for informal adjudications is the constitutional due process right plus any specifics of the statute defining the program at issue.[23]

The APA also creates a right of judicial review to challenge an agency decision on the grounds that it is "arbitrary, capricious, an abuse of discretion, or otherwise not in accordance with law; … in excess of statutory … authority, or limitations, or short of statutory right; … unsupported by substantial evidence …; or unwarranted by the facts … ."[24] There are two main categories of challenges to agency actions under the APA, substantive challenges and procedural challenges. A substantive challenge is when a party challenges an agency decision on its merits. For example, a statute says that the Department of Transportation cannot use federal funds to finance the construction of highways through public parks if a feasible and prudent alternative route exists. So a substantive challenge to a decision to construct a highway through a public park might argue that the agency wrongly concluded that there is no alternative route. In a procedural challenge, a party challenges the processes that the agency used to take an action. Because the APA requires agencies to provide the public with notice of rulemakings and a meaningful opportunity to comment, a procedural challenge might say that the agency disclosed too little information about its decision-making process for the public to meaningfully comment.

The full implications of the notice and comment rules as applied to the adoption of advanced automated decision-making technologies have not yet been defined. For example, if a matter subject to rulemaking involved the use of an AI-driven algorithm, would the algorithm or the model itself need to be put out for notice and comment? Would it be sufficient to explain how the model worked? If so, what does it mean to explain a model sufficiently to ensure that the public had the opportunity to meaningfully comment? And because training data is an important aspect of an ML model, would there need to be a notice when the training data is updated?

The workshop focused mainly on adjudications affecting individuals, such as agency decisions regarding veterans' or health benefits, but the adoption of advanced decision-making technologies can blur the line between adjudications and rulemakings. AI tools are designed to be used at scale, and thus the decision to adopt them has many of the aspects of rulemaking, while adjudicative procedures are designed to reach and review a single decision. Thus, adjudicative hearings may be an ineffective tool for identifying systemic errors in AI tools. Furthermore, while due process rules may specify that an individual has a right to a human decision maker, does the human's reliance on an AI/ML system diminish that right? Or, as AI systems become more powerful, might it be arbitrary and capricious to *not* have an AI decision maker?

---

[22] 5 U.S.C. § 551(7).

[23] See Engstrom et al., *supra* note 7, at pp. 37-39.

[24] 5 U.S.C. § 706.

### *The Equal Protection Clause and Anti-Discrimination Laws*

A third source of contestability rights are the Equal Protection Clause of the Constitution and anti-discrimination laws, which give individuals the right to challenge discriminatory actions by government agencies and the private sector. It has long been recognized that AI-based systems may replicate various kinds of prohibited bias. Federal agencies enforcing the anti-discrimination laws have made clear that those laws apply to uses of advanced technologies in employment, housing, lending, and other contexts,[25] and government decisions are also subject to challenge on the basis of discrimination across a range of protected characteristics. However, advanced decision-making technologies may pose unique questions under these principles. For example, would an affected individual be able to scrutinize an advanced system to determine that it uses factors that can be proxies for race or other protected characteristics?

### *MACHINE LEARNING AND "UNDERSTANDING"*

A confluence of technological changes over the last two decades—the rise of the Internet and the consequent availability of vast amounts of digitized data, increasing computing speed (doubling every 18 months), and statistical learning, useful for classifying data—has greatly improved the capabilities of AI/ML systems. This has enabled solutions to an enormous variety of problems ranging from language translation[26] to determining protein structure.[27] With notably mixed success, AI/ML systems have also been applied in decision-making scenarios that involve people. Facial recognition systems, used for everything from passport control to unlocking user smartphones, have been found to have a high rate of false positives and false negatives for different demographic groups.[28] Racial bias has been found in advanced automated decision-making systems ranging from those supporting health care[29] to those used in criminal justice.[30]

---

[25] Joint Statement on Enforcement Efforts Against Discrimination and Bias in Automated Systems (Apr. 25, 2023), https://www.ftc.gov/system/files/ftc_gov/pdf/EEOC-CRT-FTC-CFPB-AI-Joint-Statement%28final%29.pdf.

[26] Jeff Dean, *A decade in deep learning, and what's next* (May 18, 2021), https://blog.google/technology/ai/decade-deep-learning-and-whats-next/ [last viewed Apr. 24, 2024].

[27] John Jumper et al., *Highly accurate protein structure prediction with AlphaFold,* 596 Nature, August 26, 2021, 583.

[28] See, for example, Chapter 5 of this summary for a discussion of problems with facial recognition technology.

The workshop did not reach consensus on whether *all* AI/ML systems needed to be designed as to enable model understanding. Some researchers felt that for some applications, such as serving ads, the AI/ML system need not be understandable; others, because of bias or other issues (bias has been found in online ad generation),[31] believed otherwise. There was, however, strong agreement on the necessity of model understanding for enabling contestability of advanced automated systems making decisions about individuals. What constituted model understanding from a technical standpoint did not fully generate a consensus.

The workshop focused on how to ensure advanced automated decision-making systems were developed, procured, and used to enable contestable decisions. Successfully contesting the decision of advanced automated system requires having a sufficiently detailed understanding of how the system made its decision.

Model understanding, however, provides significant benefits beyond contestability. Understanding upon which data the system is relying to make predictions helps clarify whether the system's decision model is valid.[32] Model understanding also aids in determining whether the training data is representative of the data encountered in practice.

It has often been accepted in machine learning that in order to improve result accuracy, it is vastly more useful to increase the amount of training data as opposed to the algorithm. There is less truth in this statement than is first apparent. Much depends on the nature of the training data. Biased data that fails to fully capture the intended population in appropriate ratios will not lead to more accurate systems.

Advanced automated systems designed through a heavy reliance on data and without much emphasis on model interpretability can be confusing—and thus difficult for anyone, including experts, to understand or explain the reasoning behind the functioning of the system. A trenchant example is the ML system that incorrectly concluded that pneumonia patients with a history of asthma have a lower risk of death than those without such a history. The problem was that the model lacked knowledge that pneumonia patients with a history of asthma were quickly admitted to a hospital ICU; receiving greater care, the patients with a history of asthma had a higher survival rate than those without.[33]

---

[29] See, e.g., Ruha Benjamin, *Assessing risk, automating racism*, 366 Science, Oct 25, 2019, 421.

[30] Julia Angwin, Jeff Larson, Surya Mattu, and Lauren Kirchner, *Machine bias* (2022), Ethics of data and analytics, Auerbach Publications.

[31] Online advertising has been shown to exhibit both racial and gender bias in terms of which groups are shown particular ads. See, e.g., Muhammad Ali, Piotr Sapiezynski, Miranda Bogen, Aleksandra Koralova, Alan Mislove, and Aaron Rieke**,** *Discrimination through optimization: How Facebook's ad delivery can lead to biased outcomes* (2019), Proceedings of the ACM on Human-Computer Interaction, https://doi.org/10.1145/3359301.

[32] Correlation does not mean causation, but an ML system can confuse the two. See, e.g., Benjamin, *supra*, note 16, and Rich Caruana, Yin Lou, Johannes Gehrke, Paul Koch, Marc Sturm, and Noemie Elhadad (2015), *Intelligible models for healthcare: Predicting pneumonia risk and hospital 30-day readmission,* Proceedings of the 21st ACM SIGKDD international conference on knowledge discovery and data mining, https://dx.doi.org/10.1145/2783258.2788613.

Model understanding provides a global view of the ML system—or a large piece of it—but this view can easily otherwise be hidden due to the massive number of types of data upon which an ML system relies. By clarifying which data features the system is relying on, such understanding aids debugging. This is particularly important when the system is operating in a complex environment (e.g., driving an autonomous vehicle). This is because "novel ML system complexities ... may result in deep design flaws,"[34] and uncovering these are hard. By clarifying the logic of the system, model understanding sheds light on where the reasoning of the system is trustworthy—and where it might not be. Thus, model understanding can be useful when underlying circumstances of the modeled system abruptly change; in such shifts, the model may no longer accurately align with the real world. (One example of such a change is the drastic shift in online purchasing patterns that occurred at the start of the Covid-19 pandemic, rendering existing models of online shopping no longer valid.) Because model understanding presents some of the underlying logic of the system, it also helps in determining whether the system is ready for deployment.

Given all the advantages that model understanding presents, a natural question is why not always develop advanced automated decision-making systems so that they are understandable? The reason is the complexity of big data combined with complex models that can represent complicated patterns in data. Using hundreds of thousands or millions of data points to model the world can enable an AI/ML system to predict outcomes with extremely high accuracy. This is the case whether the system is predicting the spelling a user intended or a patient's risk for stroke. But systems employing models complex enough to represent patterns with such accuracy can come at the cost of a human's ability to understand the system (and thus also to contest it). To date, researchers have found popular machine learning techniques used in real-world applications offer a tradeoff between accuracy and model understandability, where accuracy refers to accuracy for use within the real world—and not training data.[35]

---

[33] Caruana et al. *supra* note 19.

[34] Dan Hendrycks, Nicholas Carlini, John Schulman and Jacob Steinhardt, *Unsolved problems in ML safety* (June 16, 2022), arXiv preprint arXiv:2109.13916. By clarifying the logic of the ML model, model understanding can help clarify where problems may lie.

[35] Gintare Karolina Dziugaite, Shai Ben-David, and Daniel M. Roy (2020), *Enforcing interpretability and its statistical impacts: Trade-offs between accuracy and interpretability*, arXiv preprint arXiv:2010.13764.

Currently when building a ML system, developers face a choice between creating so-called "white-box" systems whose underlying logic is humanly comprehensible, but whose details can still be complex, or "black-box" ones that may be more accurate on real-world cases but whose logic is effectively unavailable to human users of the systems.[36] Sometimes, however, it is possible to borrow from the growing literature on interpretable AI towards building inherently interpretable AI/ML decision-making systems.

### *Inherently Interpretable AI/ML Systems*

It may be the case, for example, that the features of the data naturally fall into natural clusters that are easily separated by a line (if the feature data is two dimensional) or a hyperplane (if the feature data is of higher dimension) that allow clear separation between the clusters. If so, the logic of the decision-making system is easy to explain. However, even if feature variables do not separate in such a simple way, there remain ways to develop AI/ML systems that can make predictions and are inherently interpretable by people. Currently four such ways have been developed: rule-based systems, risk-score systems, generalized additive models, and type models.

*Rule-based systems* are predictive models built through using explicit logical rules, which enables system interpretability.[37] An example of this approach is the Bayesian Rules List (BRL), which provides a set of "if X, then Y" rules from a larger set of pre-determined rules.[38] Such an approach tends to be accurate and have its logic interpretable by people.[39] BRL was

---

[36] See, e.g., Ricardo Guidotti, Anna Monreale, Salvatore Ruggieri, Franco Turini, Fosca Giannotti, and Dino Pedreschi, *A survey of methods for explaining black box models* (2018), 51 ACM computing surveys.

[37] Benjamin Letham, Cynthia Rudin, Tyler H. McCormick, and David Madigan, *Interpretable classifiers using rules and bayesian analysis: Building a better stroke prediction model* (Sep 2016), 9 Annals of Applied Statistics, 1350; Himabindu Lakkaraju, Stephen H. Bach, and Jure Leskovec, *Interpretable decision sets: A joint framework for description and prediction* (2016), Proceedings of the 22nd ACM SIGKDD international conference on knowledge discovery and data mining; Oscar Li, Hao Liu, Chaofan Chen, and Cynthia Rudin, *Deep learning for case-based reasoning through prototypes: A neural network that explains its predictions* (2018), 32 Proceedings of the AAAI Conference on Artificial Intelligence; Caruana et al., *supra* note 19.

[38] Letham et al., *supra* note 24.

[39] Letham et al., *supra* note 24, explain:

> BRL provides a new type of balance between accuracy, interpretability and computation … BRL strikes a balance … in that its solutions are not constructed in a greedy way involving splitting and pruning, yet it can solve problems at the scale required to have an impact in real problems in science or society, including modern healthcare.
>
> A major source of BRL's practical feasibility is the fact that it uses pre-mined rules, which reduces the model space to that of permutations of rules as opposed to all possible sets of splits. The complexity of the problem then depends on the number of pre-mined rules rather than on the full space of feature combinations; in a sense, this algorithm scales with the sparsity of the data set rather than the number of features. As long as the pre-mined set of rules is sufficiently expressive, used to predict stroke risk using observed medical data from 12,586 patients and based on 4,418 medical factors:

> **If** hemiplegia **and** age > 60 **then** *stroke risk* 58.9% (53.8%-63.8%)
>
> **else if** cerebrovascular disorder **then** *stroke risk* 47.8% (44.8%-50.7%)
>
> **else if** transient ischaemic attack **then** stroke risk 23.8% (19.5%-28.4%)
>
> **else if** occlusion and stenosis of carotid artery without infarction **then** stroke risk 15.8% (12.2%-19.6%)
>
> **else if** altered state of consciousness **and** age>60 then stroke risk 16.0% (12.2%-20.2%)
>
> **else if** age ≤ 70 **then** stroke risk 4.6% (3.9%-5.4%)
>
> **else** stroke risk 8.7% (7.9%-9.6%).[40]

The model was both more accurate than other ML systems developed and more comprehensible. The brevity of the rule-based system above contributes to its being convincing; it is possible for people, including domain experts, to comprehend the underlying logic. The conciseness also makes it easier to conduct troubleshooting.[41]

A second way to develop an inherently understandable AI/ML system is through *risk scores*. Features that may influence the outcome—in the case of stroke prediction, hemiplegia, age, cerebrovascular disorder, etc.; in the case of determining eligibility for a loan, has savings account with balance of at least X for at least Y years, has checking account balance of at least Z over last year, has a guarantor, etc.—carry a certain "risk" or point value.[42] By identifying the features that resulted in a decision, such a system may enable contestability if, for example, an individual can show that the input data regarding a specific feature was incorrect.[43] (Of course, it is a separate question whether the individual can challenge the selection or weighting of the features themselves.)

---

an accurate decision list can be found and, in fact, the smaller model space might improve generalization.

[40] Ibid.

[41] Ibid.

[42] Berk Ustun, Alexander Spangher, and Yang Liu, Actionable Recourse in Linear Classification. ACM Conference on Fairness, Accountability and Transparency (2019), https://doi.org/10.1145/3287560.3287566.

[43] Ibid. See Chapter 5 of this summary for constraints on the ML model, such as monotonicity, so that the response "makes sense" to an individual.

One other method to develop inherently understandable predictive AI/ML systems—generalized additive models[44]—was also briefly touched upon during the workshop.

Because inherently interpretable models can accurately make predictions while being comprehensible, they fill an important need. But many of the most widely used ML systems—examples include ChatGPT and facial recognition technologies—are not inherently interpretable; they are nonetheless used because of their perceived ability to provide highly accurate predictions.[45] Their underlying complexity—and thus intrinsic failure to produce an explanation on how the prediction was reached—is highly problematic for several reasons. First, it is an inherently human desire to understand how a decision was reached. Second, in the case of contestable decisions, understandability of the system is necessary for contestability. Third, such explanations are necessary if these systems are to be accepted; as one group of researchers explained, "Building trust is essential to increase societal acceptance of algorithmic decision-making."[46]

### *Post Hoc Explanations*

One solution to the use of non-interpretable ML models in situations where an explanation is highly sought has been to develop post hoc explanations of the reasoning of such black-box system.[47] Because the explanation must be faithful to the ML model and interpretable to the system user, it is complicated to pull off both goals. Not surprisingly, this is not always achievable.

What constitutes an explanation varies with the system user. Sending the system parameters to another researcher in the same field might suffice to show how the model is working but would fail to be useful to almost anyone else (and certainly not to a non-technically trained user!). Sending many example predictions might be a reasonable way to provide an explanation of the system's functioning, as might providing a program, a modified, shortened decision tree, the most important features influencing the predictions, or even ways to "flip" a model's prediction.[48]

---

[44] Yin Lou, Rich Caruana, Johannes Gehrke, and Giles Hooker, *Accurate intelligible models with pairwise interactions,* Proceedings of the 19th ACM SIGKDD international conference on Knowledge discovery and data mining, 623 (2013).

[45] This is despite the "hallucination" problems with ChatGPT and the known bias issues in FRT technologies.

[46] Sandra Wachter, Brent Mittelstadt, and Chris Russell, *Counterfactual explanations without opening the black box: Automated decisions and the GDPR*, 31 Harvard Journal of Law and Technology (2017).

[47] See, e.g., Marco Tulio Ribeiro, Sameer Singh, and Carlos Guestrin, *"Why should I trust you?" Explaining the predictions of any classifier*, Proceedings of the 22nd ACM SIGKDD international conference on knowledge discovery and data mining (2016), https://dl.acm.org/doi/10.1145/2939672.2939778; Scott M. Lundberg, and Su-In Lee, *A unified approach to interpreting model prediction,* 30 Advances in neural information processing systems (2017).

This last technique, *counterfactual explanations*, has become quite popular as it has the added advantage of providing an individual with recourse—a requirement under various legal regimes—if they receive an unfavorable decision. For example, if someone is turned down for a loan, then other points in the decision space are found where, if the person had had those values, such as a higher income, a longer time holding a checking account, or more money in savings account, the loan would have been approved. This "solution" of a counterfactual, however, works better in an idealized mathematical situation than in practice, an issue we will discuss shortly. Returning to when an explanation of an advanced automated system is sufficiently clear depends on the knowledge and technical sophistication of the system's user. In contestability situations, it should be assumed that the user does not have technical or domain knowledge expertise.

At a high level, approaches for post hoc explanations split into two approaches: *local explanations* explain the rationale behind individual predictions (they explain what factors impacted this particular decision), while *global explanations* focus on explaining the big picture behavior of the underlying model. The two approaches provide different benefits. Local explanations help the user understand that an individual prediction was made for the right reasons,[49] while global explanations help provide assurance that the model is behaving appropriately and thus is ready for deployment. (Note that there is not a universal understanding of what it means for an explanation to be "local," creating a problem for use of this explanation method.) In both approaches, these explanations serve to approximate the behavior of complex systems.[50]

This is not a matter, however, of having your cake and eating it. Simplifying the description of a complex ML system to provide understandability means that inevitably detail is lost. A simplified description cannot always fully capture the behavior of the underlying model. A serious problem is that the explanation may not match the way the model actually makes decisions.

There are other problems as well. Post hoc explanations are subject to "fragility"; that is, they can be manipulated to misattribute a model's output to certain features.[51] Through gaming current explanation tools, researchers have shown how to hide what features a model is actually using in making its predictions, appearing, for example, to be relying on innocuous features for predictions when the situation is otherwise.[52] Stability is a difficulty; small changes in the input can lead to large changes in the resulting post hoc explanations (this is due to the way current post hoc explanation tools work).

---

[48] Hima Lakkaraju, *ML explanability Part 3: Post hoc explanation methods* (2022), Explainability Workshop, Stanford University, https://www.youtube.coAt/watch?v=_6n8r523QP8 [last viewed Apr. 30, 2024].

[49] See, e.g., Ribeiro et al. *supra* note 34.

[50] For a more detailed explanation of local and global explanations, see, e.g., Lakkaraju, supra note 35.

[51] Dylan Slack, Sophie Hilgard, Emily Jia, Sameer Singh, and Himabindu Lakkaraju, *Fooling LIME and SHAP: Adversarial attacks on post hoc explanation methods*, Proceedings of the AAAI/ACM Conference on AI, Ethics, and Society, 180 (2020), https://dl.acm.org/doi/10.1145/3375627.3375830.

An additional issue is that there is not universal agreement on what constitutes an explanation of an ML system or how to arrive at such an explanation. Indeed, even systems that focus on the same type of "explanation" (e.g., such as providing the most influential features in the prediction) may view the problem differently and present different features. This lack of consensus makes it very difficult for a non-expert to evaluate how an ML system is making its decisions, let alone contest unfavorable ones.

### *CONTESTABILITY AND ML SYSTEMS*

In seeking to understand contestability and AI/ML decision-making systems, it is useful to first consider contestability and non-AI/ML advanced automated decision-making systems.

If the decision procedure has fixed rules, contestation is about the information the system has regarding the individual or about a misapplication of stated rules. Contesting a decision about information would involve contesting an error of fact, such as the system had the wrong information inputted ("My age is 66, not 65" or "I have been at my position for 12 years, not two!"). Contesting about rules could be that a rule was misapplied ("I was born in 1954, so I am eligible for full Social Security benefits in 2020, not 2021!") or a more complex misapplication, such as one involving due process in determining Medicare benefits.[53]

It could also be possible that the law is not fully precise, that is, there is some discretion as to which rule or process could apply to a particular situation. Because there is not an a priori agreement as to what the decision ought to be, this introduces a third type form of contesting the decision, namely objection to the procedure that was used for decision-making. An example of this would be using a procedure that is not, on its face, discriminatory, but in fact creates a disparate impact on different groups.[54]

---

[52] For example, a system might be relying on race as a feature, but the explanation fails to show this (Slack et al.).

[53] This was the case, for example, in *K.W. ex rel. D.W. v. Armstrong*, 180 F. Supp. 3d 703, 718 (D. Idaho 2016). The Idaho Department of Health and Welfare's advanced automated system had violated beneficiaries' due process rights.

[54] See, e.g., Michael Feldman, Sorelle A. Friedler, John Moeller, Carlos Scheidegger, and Suresh Venkatasubramanian, *Certifying and removing disparate impact*, Proceedings of the 21st ACM SIGKDD international conference on knowledge discovery and data mining (2015), 259.

Introducing ML data-driven processes for decision making complicates the situation. Now, ML systems do not actually "make decisions"; rather, they perform pattern recognition which is then used to make predictions. ML systems, which are more complex than the decision-making systems we have described so far, can do pattern recognition using various kinds of techniques such as unsupervised learning, supervised learning, etc. An ML system has both training data, the data used to learn the patterns, and an algorithm.

The set of training data input to the ML algorithm is not data about the individual in question, but rather a set with data on a very large number of individuals; it is possible—indeed likely—that none of these have any connection with the person in question. The fact that *the specifics of the dataset will influence the patterns that are uncovered and will be used for the predictions made by the ML system* is critical. The variability in possible training sets greatly affects the predictions the ML system will then produce.

The validity of training data is a major concern for ML systems and is a major potential target for contesting decisions based on the system. Was the ML system trained on a relevant population of interest? Is the data representative? Did the dataset omit people from a particular demographic? Was there hidden bias[55] in the training data?[56] In a study on the difficulty of finding an appropriate set of training data, industry practitioners told researchers, "You'll know if there's fairness issues if someone raises hell online."[57] It is currently the case that ML practitioners cannot access the training data they would like.[58] They instead settle for a data source they can access. All these issues open possibilities for contestation.

Once a pattern-recognition system has been developed, the next step is to develop a mechanism for predictions. This introduces additional volatility to the ML system. There is choice in which function is optimized; certain choices can lead to disparate impacts on different demographics.[59] Is the prediction target the actual target of interest? What exactly is the system attempting to learn? Even slight modifications of answers to these questions can lead the system to produce very different predictions.

---

[55] Obviously this can refer to illegal bias, e.g., disfavoring a protected class. But it can also refer to statistical bias, which in turn can lead to illegal bias or simply bad outcomes.

[56] Datasheets are a way to establish provenance of the data and create standards for its use in ML systems; see, e.g., Timnit Gebru, Jamie Morgenstern, Briana Vecchione, Jennifer Wortman Vaughan, Hanna Wallach, Hal Daumé III, and Kate Crawford, *Datasheets for dataset,* 64 Communications of the ACM (December 2021), 86.

[57] Kenneth Holstein, Jennifer Wortman Vaughan, Hal Daumé III, Miro Dudik, and Hanna Wallach, *Improving fairness in machine learning systems: What do industry practitioners need?,* Proceedings of the 2019 CHI conference on human factors in computing systems (2019), https://dl.acm.org/doi/10.1145/3290605.3300830. There is another issue lurking; what constitutes "fair" in an ML algorithm is very hard to define; see, e.g., Arvind Narayanan, *Translation tutorial: 21 fairness definitions and their politics* (2018), Proceedings of the 2018 conference on fairness, accountability, and transparency, and Rachel Courtland, *The bias detectives*, 558 Nature (June 20, 2018), 357.

[58] Holstein et al, *supra* note 44.

[59] See Narayanan *supra* note 44, Courtland supra note 44; see also Sam Corbett-Davies, Emma Pierson, Avi Feller, and Sharad Goel, *A computer program used for bail and sentencing decisions was labeled biased against Blacks. It's actually not that clear*, Washington Post (Oct 17, 2016).

ML models are often very complex. This should not be a surprise, since the very reason to seek an ML model for prediction is because of the lack of a simple transformation of the features to fully accurate predictions; indeed, there may not be such a simple transformation. But because it is exceedingly difficult to interrogate the ML model to determine how it reaches its conclusions, we are left in an unhappy state—and a highly problematic one from the point of view of contestability.

ML-based decision-making systems add several potential contestability issues. Is the training being conducted on the right set of features? Are all demographics groups that should be included in the training data present? Is everyone facing the same ML model? Is the target the relevant one? Is the post hoc decision reasoning valid? If the model is black-box, do the statistical results demonstrate disparate impacts on different demographics? The fact that real-world decisions can be difficult to model as simple mathematical functions adds yet another problem to using ML systems in automated decision-making for individuals, opening up an additional form of contestation.[60] The result: A data-driven ML model for advanced automated decision-making system may have many more contestable aspects than a simple rules-based decision model.[61]

ML systems not explicitly designed to be understandable may have post hoc explanations. A tempting form of explainability that would appear to aid in providing contestability is counterfactuals. By offering explanations of how, if some aspect of the individual's case were different, the outcome would be different, counterfactuals appear to show how the advanced automated decision-making system would change a response from negative to positive (or vice versa), that is taking a "no" for a loan to a "yes" to the loan. This apparent utility is undermined by the fact that the actions proposed may not actually be plausible ones for the data subject.[62]

In addition, because data subjects may respond to counterfactuals by changing their actions, this creates instability in the model. A counterfactual response provided at one time may no longer be valid sometime later (see Chapter 5 for further discussion of this issue).[63] While there have been efforts to model such types of recourse mathematically in order to put rigor into the system,[64] such approaches are currently not complex enough to actually capture real-world situation—and thus are not useful.

---

[60] Zachary C. Lipton, *The mythos of model interpretability: In machine learning, the concept of interpretability is both important and slippery* 16 Queue (May-June 2018), 31.

[61] These include such issues as: Was the model faithfully applied? Was the same model applied to all stakeholders?
[62] Solon Barocas, Andrew D. Selbst, and Manish Raghavan, *The hidden assumptions behind counterfactual explanations and principal reasons*, Proceedings of the 2020 conference on fairness, accountability, and transparency (2020), https://dl.acm.org/doi/10.1145/3351095.3372830.

[63] Ibid.

All this said, it is still possible to have expectations—and yes, legislation, regulations, or norms—regarding the contestability of ML-based decision-making systems. Such requirements can focus on correctness of inputs to a model, access to a model sufficient to interrogate its statistical properties (and thus learn, for example, about disparate outcomes) and disclosure of important aspects of the system (e.g., how is the training data collected, what were the inputs to the model, how were they measured, what was the cohort, what was the inclusion criteria for the cohort, what was the prediction target, what is the system attempting to learn (this is called the "learning objective"). There could be requirements for transparency of the post hoc decision logic and how predictions from the model make their way into decisions. There could be requirements about testing whether the model is suitable for different populations; this does not require looking inside the model's logic but instead involves testing the model on different demographics. And contestability could include questioning the validity of using statistical predictive modeling for certain types of problems (e.g., predictive policing).

## *PROCUREMENT CONCERNS*

The legal requirements for contestability have implications for government agencies' procurement policies. Modern organizational theory posits that large institutions should do their core competencies and outsource all else. But for an individual to effectively contest a decision made by an advanced automated system, the individual needs the ability to understand how the decision was reached.[65] As we have seen, however, data-driven ML models for advanced automated decision-making systems may obfuscate such information. Thus, is it appropriate for government agencies to outsource the development of advanced automated decision-making systems that focus on decisions regarding individuals?

A 2020 study on the use of AI systems by federal agencies observed that, "Contracting out makes more sense for police cars than for police officers."[66] Building cars is not a core competency of government, so it should outsource the manufacturing of police car to automakers. But the exercise of arrest powers and use of deadly force are, and should be, a core function of government and should not be outsourced. However, the build internally/outsource dichotomy does not work neatly for AI. By making judgements about individuals, an advanced automated system for policing that involves decision making about individuals is actually performing a role of government. But the government does not have—*and is unlikely to ever have*—the competency to build its own AI-systems the way it hires, trains, and supervises police officers. Instead, government must develop expertise in overseeing the contractors that will build and sometimes operate its advanced automated systems. Outsourcing the development of AI software without deep involvement by the federal agency involved is like contracting out police officers, an abrogation of responsibility and a recipe for disaster.

---

[64] Amir-Hossein Karimi, Gilles Barthe, Bernhard Schölkopf, and Isabel Valera, *A survey of algorithmic recourse: contrastive explanations and consequential recommendations*, 95 ACM Computing Surveys (2022), https://dl.acm.org/doi/pdf/10.1145/3527848.

[65] Indeed, the lack of adequate explanation as to why an individual's benefit had been reduced was the basis for the initial ruling in *K.W. ex rel. D.W. v. Armstrong*, in which the state of Idaho was enjoined from using its advanced automated system for calculating the number of care hours to which the individual was entitled. 180 F. Supp. 3d 703, 718 (D. Idaho 2016).

[66] Engstrom et al., *supra* note 7, at 71.

Expertise, whether on policing decisions, Medicare benefits, contestation of IRS decisions, lies within the agencies. Someone within the government agency seeking software to assist in any of these functions should know enough about both the technology and the use case to be able to ascertain whether the right training set was used, what the actual model being used was, whether it was appropriate for the task, etc.  A strong conclusion from our workshop was the supreme importance of having such expertise within government agencies.

### *CONTINUING CONCERNS*

One of the most important unresolved issues of post hoc explanations is whether these can sufficiently capture the behavior of the underlying model faithfully while also being easy to understand to a non-technical user. Another concern, to which there has not been much attention so far, is the privacy risks posed by contestability. Research has shown that techniques adopted to provide algorithmic recourse to affected individuals can often be leveraged by an adversary to leak sensitive information about the training data of the underlying model.[67] In designing contestable systems, it will be important to understand these trade-offs, and whether the ML system can provide adequate privacy protections while nonetheless enabling contestability.

---

[67] See, e.g., Martin Pawelczyk, Himabindu Lakkaraju, and Seth Neel, *On the privacy risks of algorithmic recourse,* International Conference on Artificial Intelligence and Statistics (2023), 9680.

# Chapter 3: Understanding the Stakes

Ultimately, the topic of contestability and advanced automated systems is important because of the harm that can be done to individuals by incorrect or unfair decisions. Veterans, disabled persons, unemployed workers, parents of small children, and others who are dependent upon— and legally entitled—to government assistance may face erroneous denial, termination or reduction of benefits, accusations of fraud, and even fines and claims for repayment of benefits never received. Taxpayers may be unfairly singled out for audit. Prospective home purchasers may be denied federal loan guarantees based on a faulty credit score. All are entitled to an understandable explanation and an opportunity to challenge the outcome. Before delving further into the nuts and bolts of advanced automated systems and contestability, it is valuable to see the importance of redress from the viewpoint of individuals affected by decisions of advanced automated systems.

### *CASE-BY-CASE CONTESTABILITY ON THE GROUND*

What does it mean to be on the receiving end of an automated decision? Suppose that you are dependent upon government assistance for health care, food, or housing. You receive a notice saying that your benefits are being reduced or terminated. The notice says nothing relevant about why the decision was made. The notice may be written in such vague terms or use such bureaucratic language that you have difficulty understanding it.

At that point, you might seek an attorney. If you are poor, your only option is legal aid, but you will have to get in line. In Arkansas, for instance, for every one legal aid attorney who might be able to assist you, there are 18,000 financially-eligible Arkansans. But even if you manage to secure a lawyer, the appeal process is difficult. You might have to physically go to the agency office, or perhaps you have to fax or mail the agency certain forms, and there may be a very tight timeline for that, such as 10 days from when the agency sent the notice, not necessarily from when you received it. The government is supposed to continue your benefits pending the appeal, but state agencies or the Social Security Administration often do not do that correctly.

You may have a right to an administrative hearing, but your claim file may be incomplete, and you have limited methods of discovery and difficulty obtaining or serving subpoenas. You may not know exactly what you have to prove, especially when an algorithm is involved, because you don't understand it and it hasn't been explained to you. You do not know who the agency witnesses are or what they will say. The usual rules of evidence do not apply, so often hearsay may be sufficient. It's not always clear what sort of arguments should be presented. There may be new documents introduced at the hearing that your lawyer has never seen and that haven't been shared with you before.

And then there is a phenomenon that might be called "criteria displacement." Suppose the hearing is about Medicaid care allocation. The actual criterion is "medical necessity." But the administrative law judge may not want to hear about medical necessity at all and will rule it irrelevant. So the fact that you need 2 hours to get out of bed in the morning or an hour to bathe is irrelevant. The only thing the judge wants to know is whether the inputs into the algorithm were correct at the time the assessment was done. There are 286 questions on the assessment, and only 60 of those count, so you have to show both that the nurse evaluator filled out one of the 286 wrong and that the faulty input was for one of the questions that was material. But how do you know it was actually material without understanding the algorithm?

What the judge has done is to replace the legal criterion of medical necessity (which you may be prepared to present evidence on) with the question of whether the algorithm was applied correctly. The algorithm becomes the thing it is purporting to measure. It displaces other criteria, and the judge defers to whatever the algorithm says, regardless of whether the state has validated either the mechanics of the system or the data modeling that underlies it.

Given this reality, fewer than 5 percent of appeals are successful. Usually most of those are when the person is represented by an attorney. Even if you win your case, the agency will not revise the system; at best, the agency will change only the outcome in your case, leaving the system in place.

### *SYSTEMIC CHALLENGES: A LONG AND WINDING ROAD*

If a systemic challenge is brought, it can face all the delays and hurdles associated with the most complicated commercial litigation. Consider the litigation that has been ongoing since 2012 over an automated system used in Idaho's Medicaid system. The technology at issue in Idaho is not advanced, but the case illustrates many of the barriers faced when advocates seek to systemically challenge an automated decision-making system on behalf of affected individuals.

In 2011, Idaho adopted a new program to assess an individual's need for home- and community-based services (HCBS) under Medicaid.[68] HCBS programs help disabled people perform activities of daily living in their own homes instead of in institutions. Most of the funding is federal, and the federal government can specify program requirements. But benefits are administered by states, which have been afforded significant autonomy. Under the Idaho program, a person would travel to a medical assessment center where an Independent Assessment Provider (IAP) would complete a proprietary form that scored the person's need for assistance in feeding, toileting, dressing, and other functions. The IAP would manually enter that data into a digital Budget Tool, which, in turn, automatically calculated an assigned budget amount for those reported needs based on formulas that Idaho claimed to be a "trade secret."

---

[68] The description in this paragraph and the next is drawn from Lydia X. Z. Brown, Michelle Richardson, Ridhi Shetty, Andrew Crawford, and Timothy Hoagland, *Challenging the Use of Algorithm-driven Decision-making in Benefits Determinations Affecting People with Disabilities*, CENTER FOR DEMOCRACY & TECHNOLOGY (Oct. 2020) https://cdt.org/wp-content/uploads/2020/10/2020-10-21-Challenging-the-Use-of-Algorithm-driven-Decision-making-in-Benefits-Determinations-Affecting-People-with-Disabilities.pdf. See also, *AI in Government*, Testimony of Ritchie Eppink before the Senate Committee on Homeland Security & Government Affairs (May 16, 2023) https://www.hsgac.senate.gov/wp-content/uploads/Testimony-Eppink-2023-05-16-1.pdf.

Soon after the system was implemented, Idaho's legal aid office began receiving call after call from people who discovered that the state had slashed their Medicaid benefits but had no idea why. Attorneys filed suit against the state in 2012. The case became a class action, involving thousands of Idaho residents with intellectual and developmental disabilities.

The state fought every step of the way. When plaintiffs won a preliminary ruling, *K.W. v. Armstrong*, 298 F.R.D. 479 (D. Idaho, 2014), the state appealed. The Ninth Circuit affirmed, *K.W. v. Armstrong*, 789 F.3d 962 (9th Cir. 2015), and litigation resumed. In 2016, the lower court held that Idaho's system violated constitutional due process minimums, ruled that people using Medicaid and their advocates must have access to proprietary information about the assessment instrument that provided inputs to the system, and required that Idaho ensure everyone using the Medicaid program have a "suitable representative" to help them contest system results. *K.W. v. Armstrong*, 180 F. Supp. 3d 703 (D. Idaho 2016). The two sides entered into a comprehensive and lengthy settlement agreement spelling out what Idaho was supposed to do to remedy the constitutional problems with its old system the court kept the case open to supervise implementation of the settlement. In 2023, Idaho tried to preclude class members from accessing a user's manual for the new system. When a magistrate ruled against the state, *K.W. v. Armstrong*, 683 F. Supp. 3d 1125 (D. Idaho 2023), the state asked the presiding district judge to overturn the magistrate's decision. On March 3, 2024, the district court denied the state's motion. K.W. v. Armstrong, 2024 U.S. Dist. LEXIS 39360, 2024 WL 914776 (D. Idaho, 2024).

A major focus of the litigation—and an extremely time-consuming one—was the effort of plaintiffs to uncover and analyze the hidden details of the system as the state resisted efforts to disclose any information about it. The lawyers pursuing the case initially had to fight to determine that an automated system was being used and get the formula. Then they had to fight to get the data that the formula was based on. Once they had that information, they spent tens of thousands of dollars working with multiple experts to figure out some of the flaws with the system.

In exploring why individuals' Medicaid budgets were cut sharply, it became apparent that the problems that were appearing with the automated system were actually just the tip of an iceberg that hid a whole collection of other problems. One issue was that decisions were being made about people based on factors that should have not been considered and that came to light only when program recipients became involved. One result is that the settlement agreement that resulted from the lawsuit requires the state to consult with program beneficiaries in building a new system to replace the old, flawed system.

This proved to be a key revelation about how to create a fair system—involving the people about whom the system is making decisions. Lawyers can be relied on to make sure that issues about due process are addressed, while computer scientists will concern themselves with getting the code right, but to understand how well the system will work for the people it is supposed to serve, it is necessary to involve those people in the decisions about building and using the systems, including the decision at the outset whether to use automated processes in the first place. Instead of spending resources to have experts look at the system to figure out its flaws, it can be much more effective to speak with those people who have themselves dealt with the system and found out its limitations firsthand.

> *"People with disabilities are not just a number. People with disabilities are human beings. If automated systems are making decisions about people with disabilities, people with disabilities are the primary experts on whether those systems are working fairly and correctly. People with disabilities need to be at the center of any discussions about whether automated systems will be used to make decisions about them and about whether those systems are functioning properly."*
>
> *–Kristyn Herbert,*
> *Medicaid recipient and Idaho disability rights advocate.*

# Chapter 4: Current Advanced Automated Systems in Federal and State Government

In fiscal year 2022, 20 federal agencies reported about 1,200 current and planned artificial intelligence (AI) use cases.[69] The number today is likely much higher. The Department of Health and Human Services FY 2023 inventory lists over 160 AI use cases. The current inventory for the Department of Veterans Affairs (VA) lists 127,[70] of which 40 were in an operational phase as of February 2024.[71]

The VA and other agencies stress that they are approaching the adoption of AI cautiously.[72] Indeed, we are not aware of any federal agency that is currently using or planning to use AI or other advanced techniques to autonomously make final decisions about individuals. Moreover, many current and planned uses are not rights-impacting and do not pose the contestability and due process issues addressed here. For example, the VA inventory indicates that the department is conducting a study to develop advanced AI applications in radiology, collecting unstructured data from radiology reports and using it to label radiological images at scale to train models to interpret images. In the field of medical diagnostics, it is conducting a study on applying "machine learning to the expert-labeled output of rules-based triggers, to make a two-stage diagnostic error detection process."

On the other hand, at multiple agencies AI-enabled decision support tools have been and are being integrated into benefits adjudications and other rights-affecting processes in ways that do impact decisions even if there is human review. For example, the VA's Business Transformation Portfolio automates the control of mail the VA receives in connection with benefits claims, summarizing relevant information from the file and updating claim statuses, using a variety of techniques, including neural net processing. Likewise, the Claims Profile project, which was at the feasibility study and prototype stages when reported in the department's AI inventory, will use artificial intelligence technology to build computable claim profiles from historical data of over 700 claimed disabilities, enabling adjudicators to assess new claims in the context of historical precedent. If, however, in any of these use cases human decisionmakers regularly rely upon the machine outputs, then the fact that the human makes the final decision does not obviate concerns about understandability and the other aspects of contestability.

---

[69] GAO, *Artificial Intelligence: Agencies Have Begun Implementation but Need to Complete Key Requirements* (Dec. 2023) https://www.gao.gov/assets/d24105980.pdf.

[70] VA AI Inventory, https://www.research.va.gov/naii/ai-inventory.cfm [last visited May 5, 2024].

[71] *Artificial Intelligence at VA: Exploring Its Current State and Future Possibilities*, Statement of Charles Worthington, Chief Technology Officer, Chief Artificial Intelligence Officer, Department of Veterans Affairs, before the Committee on Veterans' Affairs Subcommittee on Health, U.S. House of Representatives (Feb. 15, 2024), https://www.congress.gov/118/meeting/house/116823/witnesses/HHRG-118-VR03-Wstate-WorthingtonC-20240215.pdf.

[72] See, for example, Worthington Testimony, *supra* note 58: "VA is taking a measured approach as we begin to scale AI solutions to ensure we are adopting these powerful tools safely and in a manner that aligns with VA's mission."

*ADVANCED AUTOMATED DECISION MAKING IN GOVERNMENT TODAY*

The annual inventory of AI use cases mandated by Section 7225(a) of the Advancing American AI Act (Pub. L. 117-263) and Section 10.1(e) of Executive Order 14110 is a crucial oversight and accountability tool. However, a focus on AI alone can overlook the many ways in which agencies have long used, and continue to adopt, automated decision-making technologies that fall short of any definition of artificial intelligence but that still implicate interests of contestability. Moreover, an exclusive focus on the newness of AI can obscure the important ways in which earlier innovations in data management and analysis and earlier generations of decision support technology have laid the groundwork for AI adoption.

Consider the Social Security Administration. A 2021 report found that "the Social Security Administration (SSA) pioneered path breaking AI tools that became embedded in multiple levels of its adjudicatory process."[73] Indeed, the report concluded, SSA's Disability Program has been a "poster child for AI innovation in government." This did not happen suddenly. In the 1990s, the SSA began to systematize and digitize its core workflows through electronic systems, creating data and data infrastructure that would be highly valuable for later AI applications. As one early step, it implemented several electronic case management systems to organize its case activities and developed electronic folders to store digitized copies of evidence related to each claim. It also built out case analysis tools to structure and record staff notes and analysis about a claim's merits. In the mid-2000s, SSA created an electronic questionnaire that guided adjudicators through the steps needed to reach a disability determination. At the same time, it was developing an increasingly structured process for evaluating disability claims, based on judicial rulings reversing SSA claims decisions. The agency also identified issues related to the misinterpretation or misapplication of its policy guidance, prompting it to clarify its policies and procedures.

---

[73] Kurt Glaze, Daniel E. Ho, Gerald Ray, and Christine Tsang, Artificial Intelligence for Adjudication: The Social Security Administration and AI Governance (August 18, 2021), in Handbook on AI Governance (Oxford University Press, 2022), https://ssrn.com/abstract=3935950 [permalink: https://perma.cc/VJ7P-423Z]. The remainder of the discussion here of SSA relies heavily on the report by Glaze and colleagues. The adjudication of SSA disability claims is a good case study for automation because it involves a complicated, multi-stage analysis heavily dependent on medical data. To be eligible for SSA disability, a person must be unable to engage in any substantial gainful activity by reason of any medically determinable physical or mental impairment which can be expected to result in death or last for at least twelve months. So a claimant must show that they have "a severe and medically determinable physical or mental impairment" or combination of impairments lasting of sufficient duration and must be unable to perform any other work existing in significant numbers in the national economy, based on the claimant's residual functional capacity, age, education, and work experience.

Fast forward to today. The SSA's Quick Disability Determinations (QDD) process uses a computer-based predictive model to screen initial applications to identify cases where a favorable disability determination is highly likely and medical evidence is readily available. Another suite of decision support tools, known as Insight, analyzes decision documents as they are being written and generates alerts on potential quality issues, while also providing a variety of case-specific reference information and tools based on what Insight found in the decision's content. As Kurt Glaze and his Stanford colleagues describe it:[74]

> At the hearing level, staff use Insight to analyze draft decisions, enabling them to evaluate and react to Insight's quality feedback prior to issuance. At the appeals level, staff use Insight to analyze issued hearing decisions under their review, helping to ensure they identify and evaluate all potential quality issues prior to making a recommendation to appellate judges. Importantly, Insight is explicitly designed only as an assistive tool: It does not decide any element of a decision nor advise any specific remedy to potential quality issues.
>
> Insight's features require several AI technologies to function. First, Insight applies natural language processing (NLP) to extract information from the written decision, such as details of its findings and rationale. Insight then retrieves existing structured data about the case and claimant from workload systems (e.g., claimant claim history and biographical data, etc.). Using this more complete picture, Insight applies both rule-based and probabilistic machine learning algorithms to identify potential quality issues.

Today, as reported in its AI inventory,[75] the SSA is using multiple other AI-based tools. One model uses machine learning to identify cases most likely to have incorrect Medicare Part D subsidies and flag them for a technician's review. Another model uses machine learning to estimate the probability of resource misuse by representative payees and flag the cases for a technician to examine. A third uses machine learning techniques to identify disability cases with the greatest likelihood of medical improvement and flag them for a continuing disability review. A fourth set of models use machine learning to identify cases with greatest likelihood of disability eligibility determination error and refer them for quality review checks. The inventory includes the "anomalous iClaim predictive model," which it describes as "a machine learning model that identifies high-risk iClaims… [that] are then sent to operations for further review before additional action is taken to adjudicate the claims."

---

[74] Glaze et al., *supra* note 60.

[75] Available at https://www.ssa.gov/data/SSA-AI-Inv.csv.

However, Glaze et al. ended their study by noting that, "While SSA reports that Insight has improved quality and productivity, formal evaluations of the impact of the Insight system on accuracy and remand rates have been limited." In fact, after the Glaze study was published, the SSA Inspector General found that, despite the SSA's embrace of technology, backlogs in the disability program had increased: As of the end of FY 2023, pending initial disability claims had increased to approximately 1.13 million, double the pre-pandemic level, and pending reconsiderations had also increased to almost double the pre-pandemic level.[76] Average processing time for both workloads also increased over the same period, from 120 to 218 days for initial claims and from 109 to 213 days for reconsiderations. (On the other hand, from FY 2019 to FY 2023, SSA reduced the number of pending hearings by 44% and the average processing time for hearings also decreased from 506 to 450 days.) Moreover, SSA saw an increase in processing time from FY 2022 to FY 2023.

The Internal Revenue Service (IRS) is another agency looking to AI, building on a long-running use of automated techniques. Most prominently, perhaps, for half a century the IRS has used computers to select tax returns for audit. As the GAO reported in 1976, "Most tax returns selected at local service centers are chosen because they involve simple, readily identifiable problems usually removable by correspondence, or because they have a special feature such as an illegal deduction. Generally, the problems are identified by a computer. Sometimes district offices randomly select returns for special research programs, but generally the returns are selected because they have good audit potential. The potential is discovered by a computerized system called the Discriminant Function System (DIF)."[77] To this day, "[a]ll individual returns are computer scored under the DIF system."[78]

In September 2023, the IRS announced that "improved technology as well as Artificial Intelligence … will help IRS compliance teams better detect tax cheating, identify emerging compliance threats and improve case selection tools to avoid burdening taxpayers with needless 'no-change' audits." [79] One specific example cited was that the IRS would begin using

---

[76] SSA, Office of the Inspector General, Management Advisory Report: The Social Security Administration's Major Management and Performance Challenges During Fiscal Year 2023 (Nov. 2023) https://oig.ssa.gov/assets/uploads/022330.pdf.

[77] GAO, How the Internal Revenue Service Selects and Audits Individual Income Tax Returns (Dec 14, 1976) https://www.gao.gov/products/100316.

[78] See IRS, Internal Revenue manuals, Part 4, Examining Process at **4.1.2.6.2 (09-21-2020),** https://www.irs.gov/irm/part4/irm_04-001-002#idm140477689788048.

[79] Press Release, IRS announces sweeping effort to restore fairness to tax system with Inflation Reduction Act funding; new compliance efforts focused on increasing scrutiny on high-income, partnerships, corporations and promoters abusing tax rules on the books (Sept. 8, 2023) https://www.irs.gov/newsroom/irs-announces-sweeping-

AI to select returns of large partnerships for examination. Another example mentioned by the IRS Commissioner in a press briefing was the use of AI to sort through data reported by foreign financial institutions.[80] Another application: According to the Treasury Department's AI inventory, the Appeals Case Memorandum (ACM) leverages natural language processing capabilities to assist with extraction, consolidation, and labeling of unstructured text from ACM documents, automatic identification of key information, and processing results into a structured format.

For years, states have been increasingly turning to artificial intelligence and other automated systems to determine benefits eligibility and ferret out fraud in a variety of benefits programs, from food stamps and Medicaid to unemployment insurance.[81] For example, as of October 30th, 2023, all 58 California counties were using a single Statewide Automated Welfare System (SAWS) to support eligibility and benefits determination and public assistance case management.[82] As of March 2024, every single state, plus DC, Puerto Rico and the Virgin Islands had either completed or was in the process of developing an information technology modernization project for its unemployment insurance benefits system: 26 states had deployed their modernized system, 15 had a system under development, 10 were in the acquisition phase, and 2 were in the planning stage.[83]

### PROBLEMS WITH ACCURACY AND RELIABILITY IN GOVERNMENT AUTOMATED DECISION-MAKING AND DECISION-SUPPORT TECHNOLOGIES

As governments have adopted a variety of automated decision-making and decision support technologies, problems of accuracy and reliability have emerged, highlighting the importance of transparency, explainability, and contestability.

In 2023, Stanford researchers found that Black taxpayers were being audited at higher rates than would be expected given their share of the population. The study suggested that most of

---

effort-to-restore-fairness-to-tax-system-with-inflation-reduction-act-funding-new-compliance-efforts. See also https://www.nytimes.com/2023/09/08/us/politics/irs-deploys-artificial-intelligence-to-target-rich-partnerships.html.

[80] Janathen Allen, *How Does the IRS USE AI to Identify Tax Cheats*, IR Global (January 15, 2024), https://irglobal.com/article/how-does-the-irs-use-ai-to-identify-tax-cheats/.

[81] Michele Gilman, *States Increasingly Turn to Machine Learning and Algorithms to Detect Fraud*, U.S. News (Feb. 14, 2020) https://www.usnews.com/news/best-states/articles/2020-02-14/ai-algorithms-intended-to-detect-welfare-fraud-often-punish-the-poor-instead.

[82] See CalSAWS, *We are ONE! One Team, One System, One Goal!,* https://www.calsaws.org/ [last viewed May 29, 2024].

[83] NASWA UI Information Technology Support Center, http://itsc.org/Pages/UIITMod.aspx.

the disparity was driven by differences in audit rates among taxpayers claiming the Earned Income Tax Credit (EITC) and specifically the algorithm that was used to select returns for audit.[84] In May 2023, the IRS announced that its initial findings supported the Stanford conclusion.[85]

In 2023, the VA Office of Inspector General (OIG) found serious problems in a VA project that automated the processing of claims for hypertension (high blood pressure) claims—specifically, claims that request an increased evaluation or rating for this condition. Specifically, the project automated evidence-gathering tasks including extracting blood pressure readings and hypertension-related medication data from VA treatment records. These were compiled into a summary sheet uploaded to the veteran's electronic claims folder. Overall, the summary sheets the OIG team reviewed did not contain comprehensive blood pressure reading information that would assist claims processors in accurately deciding the claim. Testing revealed that 27 percent of all claims reviewed contained inaccurate and inconsistent determinations, resulting in inaccurate decisions on veterans' claims.[86]

Likewise, some state automation efforts have proven egregiously unreliable, producing considerable hardship for eligible claimants. A notorious example is the automated system that Michigan adopted to detect fraud in unemployment insurance claims, called MiDAS. In this case, fraud accusations were being generated algorithmically by MiDAS, with no human intervention or review. After pressure from the federal government and the filing of a federal lawsuit, and after it emerged that 64 percent of fraud claims were in the process of being reviewed or overturned on appeals in administrative court, the Michigan unemployment insurance agency finally audited the operation of the system.[87] It found that 70% of the cases in which people were assessed a fraud penalty did not actually involve fraud. From October 2013 to September 2015, MiDAS adjudicated 40,195 cases of fraud by algorithm alone, with 85 percent of those later determined to be incorrectly labeled as fraud. Even in the 22,589 cases that had some level of human interaction, subsequent review found a 44 percent error rate.

---

[84] Hadi Elzayn et al., "Measuring and Mitigating Racial Disparities in Tax Audits," Stanford University, SIEPR WP 23-02 (Jan. 2023), https://dho.stanford.edu/wp-content/uploads/IRS_Disparities.pdf.

[85] Daniel Werfel, Commissioner, Department of the Treasury, Letter to Member of the Senate, https://www.irs.gov/pub/newsroom/werfel-letter-on-audit-selection.pdf (May 15, 2023). See also https://www.washingtonpost.com/business/2024/04/12/irs-commissioner-werfel-michelle-singletary-interview/

[86] VA, Office of Inspector General, *Improvements Needed for VBA's Claims Automation Project* (Sept. 25, 2023) https://www.vaoig.gov/reports/review/improvements-needed-vbas-claims-automation-project

[87] Robert N. Charette, *Michigan's MiDAS Unemployment System: Algorithm Alchemy Created Lead, Not Gold*, IEEE Spectrum (Jan. 24, 2018) https://spectrum.ieee.org/michigans-midas-unemployment-system-algorithm-alchemy-that-created-lead-not-gold.

## THE DEVELOPING POLICY FRAMEWORK: RISK MANAGEMENT

In October 2023, President Biden issued an executive order (EO) to spur the adoption of AI in all aspects of the federal government while mitigating its substantial risks.[88] Like other executive orders, this EO does not create any right or benefit enforceable against the federal government. Nevertheless, it is intended to protect the rights of individuals by setting standards for the adoption of AI by the federal government. Moreover, it seeks to influence state and local governments (and the private sector) through guidelines, best practices, training programs, grant-making, and studies on a range of issues.

In March 2024, the federal Office of Management and Budget (OMB) issued a memorandum to the heads of all executive branch departments and agencies to further define what they must do to comply with the Biden EO.[89] The memo requires each agency (except the Department of Defense and the intelligence agencies) to individually inventory each of its AI use cases at least annually and post a public version of the inventory on the agency's website. Further, the memo requires:

> Where people interact with a service relying on the AI and are likely to be impacted by the AI, agencies must also provide reasonable and timely notice about the use of the AI and a means to directly access any public documentation about it in the use case inventory.

In addition, the memo requires agencies to consult affected communities, including underserved communities, about planned uses of AI and to solicit public feedback, where appropriate, in the design, development, and use of the AI and use such feedback to inform agency decision-making regarding the AI. The memo states that, if an agency determines based on such feedback that the use of AI in a given context would cause more harm than good, the agency should not use the AI.

In response to President Biden's October 2023 Executive Order on AI, agencies have issued a number of statements explaining how AI-based systems could violate these anti-discrimination laws:

- The Secretary of Labor published guidance for federal contractors regarding non-discrimination in hiring involving AI and other technology-based hiring systems.[90]

---

[88] Executive Order 14110, *supra* note 1.

[89] Memorandum M-24-10, *supra* note 8.

[90] U.S. Dept. of Labor, Office of Federal Contract Compliance Programs, *Artificial Intelligence and Equal Employment Opportunity for Federal Contractors* (Apr. 29, 2024) https://www.dol.gov/agencies/ofccp/ai/ai-eeo-guide.

- The Department of Housing and Urban Development (HUD) issued guidance on complying with the Fair Housing Act (FHA), the Fair Credit Reporting Act, and other relevant Federal laws when using tenant screening systems, which use data, such as criminal records, eviction records, and credit information, that can lead to discriminatory outcomes.[91]
- HUD also issued guidance addressing how the FHA, the Consumer Financial Protection Act, or the ECOA apply to the advertising of housing, credit, and other real estate-related transactions through digital platforms, including those that use algorithms to facilitate advertising delivery, as well as on best practices to avoid violations of Federal law.[92]

In April 2024, the federal Department of Health and Human Services (HHS) issued guidance on promoting the responsible use of AI in automated and algorithmic systems by state, local, tribal, and territorial governments in public benefit administration.[93] The recommendations in the plan are not mandatory and are general in nature. They are based on a concept of risk management. HHS strongly encourages agencies to implement certain specified practices to promote safe development, use, and operations of automated or algorithmic systems, such as conducting an impact assessment to determine the estimated benefit from the automated or algorithmic system as compared to its potential risks and measuring the quality and appropriateness of the data used in the system's training, testing, and prediction. The U.S. Department of Agriculture issued similar guidance for the 16 federal nutrition programs that it manages, serving a variety of populations, from infants and children to the elderly.[94]

Similar efforts are slowly underway at the state level. For example, the state of California has adopted guidelines for state use of generative AI,[95] described further in the chapter on procurement.[96]

---

[91] *Guidance on Application of the Fair Housing Act to the Screening of Applicants for Rental Housing* (Apr. 2024) https://www.hud.gov/sites/dfiles/FHEO/documents/FHEO_Guidance_on_Screening_of_Applicants_for_Rental_Housing.pdf

[92] U.S. Dept. of Housing and Urban Development, *Guidance on Application of the Fair Housing Act to the Advertising of Housing, Credit, and Other Real Estate-Related Transactions through Digital Platforms* (April 29, 2024), https://www.hud.gov/sites/dfiles/FHEO/documents/FHEO_Guidance_on_Advertising_through_Digital_Platforms.pdf.

[93] U.S. Dept. of Health and Human Services, Plan for Promoting Responsible Use of Artificial Intelligence in Automated and Algorithmic Systems by State, Local, Tribal, and Territorial Governments in Public Benefit Administration (April 29, 2024), https://www.hhs.gov/sites/default/files/public-benefits-and-ai.pdf.

[94] *Framework for State, Local, Tribal, and Territorial Use of Artificial Intelligence for Public Benefit Administration* (Apr. 29, 2024) https://www.fns.usda.gov/framework-artificial-intelligence-public-benefit.

[95] See *infra* note 123.

[96] *State of California GenAI Guidelines for Public Sector Procurement, Uses and Training* (March 2024), https://www.govops.ca.gov/wp-content/uploads/sites/11/2024/03/3.a-GenAI-Guidelines.pdf. The guidelines were issued pursuant to the state governor's Executive Order N-12-23 on Generative Artificial Intelligence (Sept. 6, 2023).

## Chapter 5: Industry Responses to AI and ML Risks

Despite problems over possible copyright infringement and hallucinations (making up "facts"), ChatGPT captured—and enraptured—the public upon its release by OpenAI in November 2022.[97] Earlier industry efforts launching ML systems in public-facing systems were not as fortuitous. Google's 2015 release of machine-learning software for labelling of Google Photos called images of Black people "gorillas,"[98] generating a public apology. In 2016, Microsoft released an ML-trained chatbot, Tay, intended to expand its vocabulary and knowledge through "conversations" with online users. Yet less than 24 hours later, Tay was pulled off from the Internet; trolls had turned Tay's use of ML technology against the chatbot and trained it to respond to users with racist and sexist communications.[99]

As these issues became evident, companies instituted a set of processes, including research, technology, and approval mechanisms, others focused largely on technical solutions. The workshop explored these approaches. Because contestability requires the ability to have a level of understanding of how an advanced automated decision-making system reaches a particular conclusion, contestability makes the use of ML solutions more complicated.

We begin this discussion by examining the ramifications from contestability, then discuss one company's approach. The goal of this chapter is not a comprehensive overview of industry's response to contestability in advanced automated decision-making systems, but instead a more limited distillation of the workshop discussion.

***COMPLEXITY OF CONTESTABLITY IN THE AL/ML CONTEXT***

Choosing to use an advanced automated decision-making system requires first recognizing that such a decision-making system is not purely a technical system, but a sociotechnical one involving interactions between technology and people. To make a fair decision about a veteran's benefits or a person's loan application, knowledge about how people function and how decisions affect human lives must be embodied in the system. While in the end, any decision-making system must determine whether a particular tax deduction is legitimate or Medicare will pay for a certain service, the development of the system must take into account that these choices involve shades of gray. Decisions about people often do.

---

[97] Warnings about the technology soon followed. Some were about false facts—"hallucinations"—in the technology's output, some were about the way the technology heavily relied upon the use of copyrighted material (see, e.g., Dan Milno and agency, *Two US lawyers fined for submitting fake court citations from ChatGPT*, Guardian (Jun. 23, 2023) and Michael M. Grynbaum and Ryan Mac, *The Times Sues OpenAI and Microsoft over A.I. Use of Copyrighted Work,* N.Y. Times (Dec. 27, 2023).) Lawyers who used the algorithm to help prepare briefs did so at their peril (Milno and agency). Other concerns included use of ChatGPT in academic settings, including the use by students to write their papers.

[98] Conor Dougherty, *Google Photos Mistakenly Labels Black People "Gorillas,"* N.Y. Times (Jul. 1, 2015).

[99] Peter Lee, *Learning from Tay's introduction*, Official Microsoft blog (May 25, 2016), https://blogs.microsoft.com/blog/2016/03/25/learning-tays-introduction/.

As sociotechnical structures, advanced automated decision-making technologies are systems in which the interactions of technology and people using it—both the operators and the stakeholders—"cannot be separated."[100] As such, an advanced automated decision-making system should incorporate the human elements from the design phase on. This includes the individuals about whom the decisions are made, the people who interact with the decision-making system, and the legal and policy choices that inform the decisions being made. As if this were not already sufficiently complex, the fact is that there is no one-size-fits-all to ensuring that advanced automated decision-making systems are genuinely contestable: Contestability is highly use-case dependent. One example of this is the problems that arise with ML-trained facial recognition technology, whose false positive and false negative rates can cause problems that range from minor annoyances (e.g., Apple's FaceID not working while wearing a medical mask) to very serious ones, such as those involving wrongful arrest.[101]

Ensuring fairness and enabling contestability in an ML-based decision-making system requires first understanding who the affected populations might be and then ensuring their participation during the decision of whether to deploy an advanced automated decision-making technology.

---

[100] Oxford Reference, *Overview: socio-technical system,* https://www.oxfordreference.com/display/10.1093/oi/authority.20110803100515814 [last viewed May 3, 2024].

[101] ML-trained facial recognition technology (FRT) has varying rates of false positives and false negatives. Systems developed in Western countries, which tend to be trained on white faces, have high false positive match rates for certain demographic groups, including women, older people, and "people of East Asian, South Asian, and African descent" (National Academies of Sciences, Engineering, and Medicine, *Facial Recognition Technology: Current Capabilities, Future Prospects, and Governance*, Washington, DC: The National Academies Press, 2015). By contrast, FRT developed in China does well on East Asian subjects and poorly on white faces (Ibid).

> The impact on individuals of false positives and false negatives caused through use of FRT varies greatly. FRT is used by the U.S. Department of Homeland Security for passenger security screening at some airports. False negatives at airport security screening usually results in a retry and short delay for the passenger. FRT use by police can result in far more problematic situations, especially in the case of false positives. Six well-documented cases of wrongful arrest and detention of Black individuals arose from false positives; in some cases, the detainment lasted multiple days (National Academies of Sciences, Engineering, and Medicine, 2). According to a National Academies study, six is likely to be but a small portion of such arrests of Black people based on false identifications by FRT. Other examples of failure of FRT include systems' inability to recognize people undergoing gender transformation surgery. False negatives in an Uber driver identification system resulted in multiple failures to recognize employees, leading to lost wages until the problem was resolved (Jaden Urbi., *Some transgender drivers are being kicked off Uber's app,* CNBC (Aug 8, 2018). https://www.cnbc.com/2018/08/08/transgender-uber-driver-suspended-tech-oversight-facial-recognition.html.

Depending on age and other circumstances, this may also include their representatives. Participation does not end at the resolution on whether to deploy an advanced automated system. If the decision is to proceed, then the affected populations and representatives should also participate during product development, procurement, testing, deployment, and updates.

This is not the only non-technical expertise needed during the design, development, testing, deployment, and updates of advanced automated decision-making systems. Technologists are always willing to throw more data at the problem—this is known to improve performance—but developing contestable advanced automated decision-making systems requires understanding the context from which the data arose and in which the systems will be used. Such understanding can illuminate important issues regarding the accuracy of the data of which technologists may be unaware; this could be due to how the data was collected, underlying societal bias at the time the historical data was collected, etc.

One trenchant example is training data for a program evaluating health profiles of patients to determine whether they should be enrolled in a "high-risk management program." The predictive determination used historical data of health costs for patient to make the determination of risk. But because this historic data was based on decisions that had Black patients relegated to using poorer facilities, the health needs of current Black patients were underestimated. Obermeyer et al. concluded that if the predictive tool had considered actual health-care needs of Black patients—as opposed to using the proxy of health-care costs—the number of Black patients receiving a recommendation for being part of the high-risk group would have doubled.[102] Determining the right "label"—on which type of data the program should be trained—is critical. That is because, "labels are often measured with errors that reflect structural inequalities."[103]

Perspective matters. For contestability, this means including sociologists, policy experts, and lawyers, including litigators with experience in contestability cases, from the start of the process of designing advanced automated decision-making systems; note that this includes in the decision about whether to deploy an advanced automated decision-making system for a particular situation in the first place. People with a perspective of the bigger social picture are crucial for ensuring the predictions of an advanced automated decision-making system appropriately serves the population about whom decisions are being made.

---

[102] See, e.g., Ziad Obermeyer, Brian Powers, Christine Vogeli, and Sendhil Mullainathan. *Dissecting racial bias in an algorithm used to manage the health of populations*, 366 Science (Oct. 25, 2019) and Benjamin, *supra* note 16.

[103] They went on to say, "Within the health sector, using mortality or readmission rates to measure hospital performance penalizes those serving poor or non-White populations ... Outside of the health arena, credit-scoring algorithms predict outcomes related to income, thus incorporating disparities in employment and salary ... Policing algorithms predict measured crime, which also reflects increased scrutiny of some groups." Obermeyer et al., *supra* note 89.

As noted, the most critical people to involve in the design, development, and initial decision to employ an advanced automated decision-making system remain the users directly affected by the decisions being made. They are the ones who understand the impact of the systems in a way no one else does. Others who should be consulted and involved through every step of design, development, procurement, use, and system updates include the others deeply involved in the system's use: operators, end users, and decision makers.[104]

Contestability means that the individual should be able to learn which factors led to their unfavorable decision—and thus how to turn the situation around. Yet current ML techniques can lead to responses in which the reasons a decision turned out unfavorably for an individual can be confounding. The individual wants to know counterfactuals. If I made $10,000 more a year, could I have gotten the loan? If I had said my knee pain was an "8" every time I stood up, would I have been approved for surgery and physical therapy? In practice, current ML techniques not specifically designed to be understandable can sometimes provide problematic and inexplicable responses.

Consider the example of an individual's loan application being rejected. On querying, the individual might learn this occurred because their income was too low. Does that mean that if their annual income rises by $10,000, they will be approved for the loan? If yes, one would also expect that if their annual income rises by $10,500, they would also be approved. But ML techniques do not currently guarantee "monotonic" solutions; that is, a rise of annual income by $10,000 might lead to a "yes" for this applicant's loan, but an annual income rise of a few dollars more than that might not. The raw way of using machine learning—relying solely on data instead of the logic of decision making—can lead to situations that make no sense even if the data appears to indicate that is the right decision.

To actually make sense—and thus be useful for contestability—the mathematical relationships should be monotonic—that is, not exhibiting behavior of jumping back-and-forth between "yes" and "no" responses (e.g., not doing "yes" for an individual's loan at an income of $50,000, "no" at an income of $50,500, "yes" at an income of $51,000)—and somewhat smooth—not exhibiting sudden, sharp discontinuities when viewed as a graph (e.g., not having surprisingly sharp increases in loan interest rates).

The need to protect the privacy of data of individuals further complicates government use of advanced automated decision-making systems. The first issue is simply securing the data; many government agencies handle highly sensitive data (e.g., the IRS and Security and Exchange Commission handle sensitive personal and business financial data, while the Department of Veterans Affairs and the Social Security administration handle sensitive health data). The more often such data is accessed, e.g., in use for training ML models, the more chance that this information might be exposed. But ML models designed to enable contestability create an additional risk. Because models retain information about the training data, it is possible that sensitive information about individuals that was used for training purposes can leak during contestation.[105] Advanced automated systems must be designed to ensure such leakage cannot occur.

---

[104] National Institute of Standards and Technology, *Artificial Intelligence Risk Management Framework (AI RMF 1.0)*, Measure 2.8 (2023), https://airc.nist.gov/AI_RMF_Knowledge_Base/Playbook.

These varied complexities, which must be addressed if ML technologies are used in contestable decision-making systems, slow the development, procurement, testing, deployment, and use of advanced automated systems. Addressing them is made more difficult by the rapidity of change of AI/ML technologies. It is something of a cliché to mention the societal disruption caused by development of technology, but the development of AI/ML is far more rapid and disruptive than previous shifts. The disruptions of the last two decades—digitization and use of the Internet for many types of activities—occurred over a period of years. By contrast, AI/ML technologies can substantively transform operations over a matter of weeks. Because designing contestability into systems requires consultations with many types of stakeholders, the rapid evolution of AI/ML systems creates additional strain in ensuring the processes to make advanced automated decision-making systems genuinely contestable occur.

With this understanding of the complexities of ensuring contestability in advanced automated decision-making systems, we turn to examining how various companies are handling the complexities posed by the need for contestability in advanced automated decision-making systems.

*ONE COMPANY'S APPROACH TO RESPONSIBLE AI IN ADVANCED AUTOMATED SYSTEMS*

There are multiple players involved in building advanced automated decision-making systems, some large, some small. We examined Microsoft's approach to doing so.

Microsoft is a technology developer, building technologies and systems for a variety of users, including other companies. It is also a platform provider, with other technology companies building applications on top of the models and platforms hosted by Microsoft. The company takes its role as a technology leader seriously. After the Tay debacle in 2016, Microsoft put significant effort into ensuring that AI/ML products developed by the company were designed to be responsible.[106]

---

[105] See, e.g., Pawelczyk *supra* note 54.

[106] Microsoft, *Responsible and Trusted AI* (July 28, 2023), https://learn.microsoft.com/en-us/azure/cloud-adoption-framework/innovate/best-practices/trusted-ai

The company started with a cross-company initiative, A*I Ethics and Effects in Engineering and Research (AETHER),* in 2017; this was led by researchers and engineers at Microsoft Research.[107] In a company as large as Microsoft—and with as many different products as the company has—understanding where AI technology is heading is critical for developing principles to guide responsible development, hence the central role that Microsoft Research played in the company's first step in responsible AI. In 2018, the company adopted AI principles developed by AETHER and in the following year, created an Office of Responsible AI to operationalize the principles and coordinate AI governance across the company.[108] In 2022, the company strengthened the principles, which govern how the company develops AI/ML products.[109]

The six principles are fairness, reliability and safety, privacy and security, inclusiveness, transparency, and accountability.[110] These are instantiated through the following product development requirements:[111]

1. Microsoft AI systems are assessed using Impact Assessments.[112]
2. Microsoft AI systems are reviewed to identify systems that may have a significant adverse impact on people, organizations, and society, and additional oversight and requirements are applied to those systems.
3. Microsoft AI systems are fit for purpose in the sense that they provide valid solutions for the problems they are designed to solve.
4. Microsoft AI systems are subject to appropriate data governance and management practices.
5. Microsoft AI systems include capabilities that support informed human oversight and control.
6. Microsoft AI systems that inform decision making by or about people are designed to support stakeholder needs for intelligibility of system behavior.[113]

---

[107] Aether, Microsoft, *Advancing AI trustworthiness: Updates on responsible AI research* (Feb. 1, 2022), https://www.microsoft.com/en-us/research/blog/advancing-ai-trustworthiness-updates-on-responsible-ai-research/ [last viewed April 20, 2024].

[108] Brad Smith., Meeting the AI moment: advancing the future through responsible AI, Microsoft on the Issues (blog), (Feb. 2. 2022), https://blogs.microsoft.com/on-the-issues/2023/02/02/responsible-ai-chatgpt-artificial-intelligence/ [last viewed April 20, 2023].

[109] Microsoft, *Microsoft Responsible AI Standard*, https://www.microsoft.com/en-us/ai/principles-and-approach [last viewed April 20, 2024].

[110] Microsoft, *Empowering Responsible AI Practices, Principles*, https://www.microsoft.com/en-us/ai/responsible-ai [last viewed May 24, 2024].

[111] Ibid.

[112] Such reviews occur at least annually.

[113] Ibid.

While the principles all have impact on ensuring contestability in AI/ML decision-making systems about people, two—Principles 2 and 6—are particularly important for doing so. Principle 6 requires that the system has to be designed so that stakeholders, defined as those who will use outputs of the system to make decisions or who are subject to decisions of the system, "can understand the system's intended uses" and can effectively interpret the system's behavior.[114] The system must be evaluated to determine whether these requirements are upheld.[115]

These requirements documents are public, thus enabling not just Microsoft's customers, including those that build applications on top of the company's services, but others to benefit from its careful work on building genuinely contestable ML systems.

There are, of course, questions about how well this system works in practice. Engineers tend to view requirements that stand between them and shipping product as a problem to be routed around. Providing engineers with easy-to-use tools for developing responsible AI can mitigate this issue. Microsoft went one step further than that, not only developing tools for error analysis, interpretability, fairness assessments, and mitigations as well as making them permanent parts of the AI platform inside the company, but also open sourcing them.[116]

Because Microsoft keeps certain information confidential, it is hard to know if the right domain expertise and specialization is sufficiently brought to decisions on whether an AI/ML system should be developed for a particular application and how the systems are being developed and designed. Furthermore, technologists—and Microsoft is a technology company—tend to focus more on such issues as model accuracy in a given context and less on user experience, how people will use these systems—or on how the systems are evaluated with real-world users.

### *COMPLEXITY OF DETERMINING USE OF AI/ML WITHIN SYSTEMS*

A common question that college faculty ask students is whether the students have used an AI or ML system within the last day. Many will answer "no." Then the faculty member will ask how many of them have used Google search. Almost all will have—and Google uses AI/ML systems in multiple different ways to process search queries.[117]

---

[114] Ibid, Requirement T1.2, p. 9.

[115] Ibid.

[116] Microsoft, *Responsible AI Toolbox*. https://github.com/microsoft/responsible-ai-toolbox [last viewed April 20, 2024].

[117] Barry Schwartz, *How Google Uses Artificial Intelligence in Google Search*, Search Engine Land (Feb. 3, 2022), https://searchengineland.com/how-google-uses-artificial-intelligence-in-google-search-379746 [last viewed Apr. 21, 2022].

The point of this anecdote is not that students fail to understand what technology they are using; rather, it is that figuring out whether a complex automated system includes any AI/ML components is increasingly difficult for anyone to determine. This is the case for complex government solutions, which often outsource system components. Loans guaranteed by the Department of Veterans Affairs (VA), for example, may use scores developed by the Fair Isaac Corporation (FICO) for determining loan eligibility. FICO, which is relied upon by many lenders for credit scoring, uses explainable ML systems for determining credit risk,[118] but the 2023 compilation of applications of AI/ML solutions by U.S. government agencies does not include a VA listing for loan applications.

---

[118] As of 2018, the company was using explainable AI/ML models; see Gerald Fahner, *Developing Transparent Credit Risk Scorecards More Effectively: An Explainable Artificial Intelligence Approach,* Data Analytics, 17 (2018).

# Chapter 6: Clarifying the Challenge of AI/ML to Contestability: Lessons from the Census Bureau's Experience with Differential Privacy

A key to successfully introducing new technology for advanced automated decision-making technologies is gaining the trust of the people who will use the new system. This requires explaining to the people affected what the technology is and what its introduction means for them. In 2020, the Census Bureau faced exactly such a challenge with the use of differential privacy[119] in publishing census data. The Bureau's experience in presenting that complex technology to a wide array of stakeholders provides valuable lessons to agencies planning on using advanced automated systems in contestable situations. During the workshop, Michael Hawes, Senior Statistician for Scientific Communication in the U.S. Census Bureau's Research and Methodology Directorate, briefed on lessons learned from the Census Bureau's experience.

### BACKGROUND

The U.S. Census Bureau does very impressive work in terms of collecting data and providing statistical information about the US population and its economy. This includes the head count for the decennial census, but also the provision of much more granular data. The information provided is highly reliable and highly trusted by its users.

Its uses and users are myriad. City officials responsible for road design and laying sewer lines seek to know how many people are really in a community. Is there sufficient provision for them? How fast is the community growing? Businesses planning expansion or utilities providing services want to know much of the same information. Academics—sociologists, economists, and others—want to study social and economic trends and behaviors. All turn to the Census Bureau as a source of reliable information.

Census data is also critical to the function of government. The allocation of billions of dollars in federal funds to localities relies on having accurate population counts. The premise of one person, one vote apportionment turns on knowing accurate counts. It is no exaggeration to say that control of the government depends on the accuracy of this data.[120]

---

[119] Cynthia Dwork, Frank McSherry, Kobi Nissim, and Adam Smith, *Calibrating noise to sensitivity in private data analysis*, Theory of Cryptography: Third Theory of Cryptography Conference, 265 (2006).

[120] The collection of census data for apportioning seats in the House of Representatives is mandated by the Constitution (U.S. Const. Art I §2 cl. 3).

Even while publishing information about the activities of people within the nation, the Census Bureau is strongly committed to protecting the privacy of the data it collects about individuals. The law requires this,[121] but it is also the case that the agency could not do its job without carefully attending to the confidentiality of the data with which it is entrusted; otherwise, individuals would be loath to provide accurate information about sensitive matters to the Census.

Privacy of an individual's data is based upon ensuring that the data is not reidentifiable, that is, it should not be possible to learn information about an individual from the aggregated statistics that the Census Bureau releases. This is a tricky issue. Statisticians know that anytime a piece of information or any statistic is released that is either from or derived from a confidential data source, releasing it reveals or leaks a tiny bit of confidential information in the process. Usually what will leak will be a quite unimportant piece of information. So statistical agencies around the world square the circle by applying a variety of statistical techniques to these data to introduce uncertainty into those results. Doing so mitigates or manages how much confidential information is leaked or revealed in the statistical products that the agencies produce.

In the mid-2000s, computer science researchers showed how that promise could be nullified by the vast amount of publicly available aggregated data; the use of this data allowed the revelation of private information about individuals. The confidentiality promises of the Census Bureau no longer held.[122] Then Chief Scientist and Associate Director for Research and Methodology of the Census Bureau, John Abowd, described this situation as "the death knell for traditional data publication systems from confidential sources."[123] The techniques that the Census Bureau had been employing could not protect against this. However, use of a new technology, differential privacy, could.[124]

---

[121] 13 U.S.C. §9.

[122] John M. Abowd, Tamara Adams, Robert Ashmead, David Darais, Sourya Dey, Simson L. Garfinkel, Nathan Goldschlag et al., *The 2010 Census Confidentiality Protections Failed, Here's How and Why*, No. w31995. National Bureau of Economic Research (2023).

[123] John Abowd, *Staring-down the database reconstruction theorem,* Talk at Joint Statistical Meetings (2018), https://www.census.gov/content/dam/Census/newsroom/press-kits/2018/jsm/jsm-presentation-database-reconstruction.pdf.

[124] Dwork, *supra* note 106. See also Daniel L. Oberski and Frauke Kreuter, *Differential privacy and social science: An urgent puzzle*, 21 Harvard Data Science Review, (2020).

## DIFFERENTIAL PRIVACY

To limit how much confidential information is revealed when statistical products are published, organizations can apply a variety of statistical techniques to these data to introduce uncertainty into those results. Data can be suppressed or redacted, coarsened through aggregation or rounding, or noise or uncertainty can be injected directly into the data. Such methods "blur" the underlying data enough that it becomes much harder to infer the original data items from the statistical information that is released.

Differential privacy provides a mathematical framework that enables statisticians to assess an individual's potential contribution to a statistic. That assessment then allows determination of how much noise needs to be added to the statistic to protect—"hide"—the individual contributions. Differential privacy is a complex technology that the Census Bureau used for releasing information about the 2020 census data along with an added wrinkle to assure the integrity of the census data. After using differential privacy to inject noise into the data on which statistics were based, the agency's statisticians post-processed the noise-infused data by imposing certain reasonableness constraints on the data, such as ensuring that subcategories correctly added up to their parent category. For example, the number of males in a particular geographic region and the number of females in that region must add up to the total number of people in that region,[125] the number of individuals in an individual census block plus the individuals in all of the census blocks around it must add up to the number of individuals in the corresponding block group or tract or county or whatever the census blocks combine to create, etc. The algorithm to do this is technically complex.

Use of differential privacy created a complex situation for the Census Bureau. While the technique was mathematically sound, its computations and proof were not easily accessible to most of those who used Census Bureau products. At issue was public trust in what the Census Bureau was doing. The agency needed to convince its stakeholders that the Census Bureau's statistical analyses could still be trusted.[126]

---

[125] The 2020 questionnaire did not include a choice of non-binary or "Other"; see https://www2.census.gov/programs-surveys/decennial/2020/technical-documentation/questionnaires-and-instructions/questionnaires/2020-informational-questionnaire-english_DI-Q1.pdf.

[126] Unhappiness with the Census Bureau's decision to use differential privacy was so great that the state of Alabama filed suit against the Bureau. The suit was dismissed without reaching a decision on the merits; see *Alabama v.United States Department of Commerce*, 546 F.Sup. 3d 1057 (2021).

***COMMUNICATING WITH STAKEHOLDERS***

As the Census Bureau proceeded towards conducting the 2020 census, many stakeholders were concerned about its new statistical direction. Congress wanted assurance of Census Bureau's ability to produce useful data. Federal and state policy makers wanted assurance that the data was sufficiently trustworthy to be used to allocate approximately $3 trillion in funding each year through funding formulas based on the data. People who decide on the boundaries of voting districts, including the districts for the House of Representatives, wanted to know that the data being produced was accurate enough to use for producing balanced districts that would conform with the Voting Rights Act. Each of the many stakeholder groups needed to be able to trust the data for its particular use. The Census Bureau realized that its different groups needed different communications on the technological changes in order to trust the output of the agency.

The Census Bureau focused on convincing its various stakeholders that the data being produced could be trusted (and, by extension, that the Census Bureau itself could be trusted). What the Bureau communicated depended on the audience. Some of the agency's messaging focused on the need to change its protection mechanisms, and some focused on the validity of the resulting data. Other messages focused on the mechanism itself so that more technically oriented stakeholders could have faith and trust in how the Census Bureau was protecting privacy or how they were injecting noise into the data.

The agency found it had to develop multi-layered messaging at different levels of technical complexity to address the different groups, targeting the messaging for each of the stakeholder groups so that the communicator focused on what was of interest to that particular stakeholder group. In short, "explanation" will mean different things for different groups. None of the stakeholder groups needed to understand the technology sufficiently to become differential privacy implementors; they just needed to be able to understand their components sufficiently to comprehend its implications. That is, a communication strategy should focus on those pieces that any particular stakeholder group needed to understand, tailoring the message to the group with the appropriate amount and level of detail.

The agency found that what worked best was having a team of people who understood the technology well enough to communicate about it but who also had the communication skills to manage stakeholder expectations and engender trust. Bandwidth meant that placing technology experts on communication tasks (and consequently removing them from implementing the technology) was a mistake; these experts were needed most in technical development. And so the agency kept its technical experts implementing the technology and placed people who understand the technology at a "good-enough" level and who were good at communications working with stakeholders.

The Census Bureau effort also worked on educating stakeholders on the new technology. Differential privacy has a "privacy budget"; this effectively allows making a choice between which data will be presented with more noise (and hence with more privacy protection) and which data will be presented more accurately.[127] This tradeoff is a necessary aspect of ensuring the privacy guarantees of differential privacy. To make these choices, Census Bureau met with multiple stakeholders to determine in which uses of census data precision was most critical.[128] To obtain well-informed responses to this question, the agency first needed to educate the stakeholders in what the mathematical tradeoff question was actually asking. Once the stakeholders understood that, they could provide appropriate answers to the Census Bureau's question.[129] The Bureau could then tweak the differential privacy system so that it would provide statistical summaries to satisfy different user needs.

In short, there were new communications challenges that needed to be met, including deciding on the appropriate degree of complexity for the communications, determining what aspects of the technology people were focusing on, and how to create this multi-layered messaging. Agencies working with new advanced automated decision-making technologies are likely to face similar sorts of challenges to those the Census Bureau faced with differential privacy.

---

[127] This tradeoff is less an issue with large populations but is especially important for smaller populations (Michael B. Hawes, *Implementing differential privacy: Seven lessons from the 2020 United States Census*, 2 Harvard Data Science Review, 4, (2020)).

[128] The Census Bureau was wrestling with questions of the sort, "Should more of the privacy-loss budget be expended on statistics that allow municipalities to know where to build hospitals and schools, or should it be spent on benchmark statistics that serve as the sampling frame and survey weights for demographic and health care surveys throughout the decade?" Ibid.

[129] Ibid.

# Chapter 7: Procurement Issues

As government agencies move to include more advanced automated technologies in their decision-making or decision-support systems, the procurement process—the nuts and bolts of government contracting—will play a major role in determining whether or not the outputs of such systems are meaningfully contestable. Workshop participants agreed that contestability cannot be added after a system is designed; it must be built in from the beginning. The goal of contestability will be achieved only if procurement specialists work with technology experts and those who will be using any system from the outset of a project. In the end, procurement officials must make sure that the contractual requirements for the system require understandability, transparency, and contestability as mandatory system features.

The rules around government procurement are complex, and the diversity of issues posed by AI adds to the challenge. Government agencies can choose a variety of means to acquire AI-enabled resources: in-house development, contracted development, licensing of AI-enabled software, and use of AI-enabled services. Generally, though, government agencies will not be building advanced automated systems themselves but rather will acquire them as products or services.  However, as noted above in Chapter 2, buying an automated decision-making system is not like buying a police car. Instead, buying a decision-making system is in many ways like training police officers: It involves non-delegable policy decisions that go the heart of how a given government agency or program operates. Therefore, government procurement officers and others inside agencies will have to be much more intimately involved in controlling the myriad choices associated with building advanced decision-making systems.

The system for any given program will often involve multiple technologies. For example, a single system might include robotic process automation, intelligent character recognition, optical character recognition, and some form of generative AI, all of which must work together—and flaws in any one of which could produce inaccurate or unreliable outputs. As a system is designed, different vendors may be producing different components. So as much complexity as there is in building advanced automated systems, there may be even more complexity in the mixing of various technological elements to create a single system. All these considerations reinforce the importance of designing for contestability from the earliest stages of a project's conceptualization.

Contestability is just one factor that needs to be considered when procuring automated systems. Other factors include bias, privacy, data rights, and security. But many of the actions undertaken to address those concerns should be leveraged to also address contestability. For example, as government officials involved in the design of a system address bias, their inquiries into the training data and their efforts to understand the resulting models should also contribute to the process of contestability by design.

Workshop participants noted that current procurement processes may have a number of weaknesses when it comes to ensuring contestability. For instance, the office procuring the technology may be separate from the agency that will be using the technology, and there may be little collaboration between the two agencies as the system is being developed. Sometimes the agency that will be using the technology lacks the technical expertise to ask the right questions about the technology. This furthers highlight the importance of assembling diverse teams that include procurement experts, technology experts, security experts, privacy experts, and individuals who will be affected by the system (or their representatives).

In some cases, such as in the Medicaid and SNAP programs, the federal government has oversight authority over the states, in particular, oversight concerning the use of federal funds to procure advanced automated systems for use in those benefit programs. This federal oversight can be vital when the states do not have the expertise to make reasonable decisions about the technologies.

It is important for the government to address intellectual property (IP) issues when developing and implementing advanced automated systems. This is a particular concern with respect to contestability, since contesting an outcome may require examining the details of the algorithm or the AI model. Yet too often developers claim that those details are protected intellectual property. Agencies should either insist on owning IP rights to the automated technologies they are using or else having a license that allows them to reveal relevant details about the technology when explaining a decision. So far this is rarely done.[130]

### *PROCUREMENT GUIDANCE TO DATE DOES NOT FULLY ADDRESS CONTESTABILITY*

The Biden administration has recognized the crucial importance of procurement processes in realizing its goals of safe, secure, and trustworthy AI, although it has not specifically called out the value and requirement of contestability.[131] Nevertheless, the foundation for contestability by design was laid by EO 14110, which directed the Office of Management and Budget (OMB) to issue guidance to agencies on the development and use of AI, including application of mandatory minimum risk-management practices to procured AI; independent evaluation of vendors' claims concerning both the effectiveness and risk mitigation of their AI offerings;

---

[130] A failure to deal with IP issues can also result in vendor lock: If the ownership of the IP was not clearly defined in the contract the agency signed with the vendor, when it comes time to recompete, the agency may find that it must either start from square one in developing a new model with a different vendor or else stick with the original vendor.

[131] The administration's Blueprint for an AI Bill of Rights, https://www.whitehouse.gov/ostp/ai-bill-of-rights/ ***[permalink: ] did highlight the importance of redress, although it did not speak to the role of the procurement process in ensuring the realization of that right:

> You should have access to timely human consideration and remedy by a fallback and escalation process if an automated system fails, it produces an error, or you would like to appeal or contest its impacts on you. Human consideration and fallback should be accessible, equitable, effective, maintained, accompanied by appropriate operator training, and should not impose an unreasonable burden on the public. Automated systems with an intended use within sensitive domains, including, but not limited to, criminal justice, employment, education, and health, should … incorporate human consideration for adverse or high-risk decisions.

documentation and oversight of procured AI; and provision of incentives for the continuous improvement of procured AI.[132]

Pursuant to the EO, in March 2024, OMB issued a memorandum titled Advancing Governance, Innovation, and Risk Management for Agency Use of Artificial Intelligence.[133] Consistent with the administration's overall approach to AI, it requires agencies to follow a risk management approach. Unfortunately, while its provisions address notice and human intervention, it does not specifically state that contestability should be a design criterion for rights-affecting systems.

Under the memo, no later than December 1, 2024 and on an ongoing basis *while using* new or existing rights-impacting AI, agencies must ensure these practices are followed for the AI:

**Provide additional human oversight, intervention, and accountability as part of decisions or actions that could result in a significant impact on rights or safety.** Agencies must assess their rights-impacting and safety-impacting uses of AI to identify any decisions or actions in which the AI is not permitted to act without additional human oversight, intervention, and accountability. [Footnote omitted.]

**Provide public notice and plain-language documentation.** … Where people interact with a service relying on the AI and are likely to be impacted by the AI, agencies must also provide reasonable and timely notice about the use of the AI and a means to directly access any public documentation about it in the use case inventory. [Footnote omitted.]

In addition, no later than December 1, 2024, agencies must also follow these minimum practices before *initiating* use of new or existing rights-impacting AI:

**Notify negatively affected individuals.** Consistent with applicable law and governmentwide guidance, agencies must notify individuals when use of the AI results in an adverse decision or action that specifically concerns them, such as the denial of benefits or deeming a transaction fraudulent. Agencies should consider the timing of their notice and when it is appropriate to provide notice in multiple languages and through alternative formats and channels, depending on the context of the AI's use. The notice must also include a clear and accessible means of contacting the agency and, where applicable, provide information to the individual on their right to appeal. Agencies must also abide by any existing obligations to provide explanations for such decisions and actions. [Footnotes omitted.]

**Maintain human consideration and remedy processes.** Where practicable and consistent with applicable law and governmentwide guidance, agencies must provide timely human consideration and potential remedy, if appropriate, to the use of the AI via a fallback and escalation system in the event that an impacted individual would like to appeal or contest the AI's negative impacts on them. Agencies that already maintain an appeal or secondary human review process for adverse actions, or for agency officials' substantive or procedural errors, can leverage and expand such processes, as appropriate, or establish new processes to meet this requirement. These remedy processes should not place unnecessary burden on the impacted individual, and agencies should follow OMB guidance on calculating administrative burden. Whenever agencies are unable to provide an opportunity for an individual to appeal due to law, governmentwide guidance, or impracticability, they must create appropriate alternative mechanisms for human oversight of the AI. [Footnote omitted.]

---

[132] Executive Order 14110, *supra* note 1.

[133] Memorandum M-24-10, *supra* note 8.

**Maintain options to opt-out for AI-enabled decisions.** Agencies must provide and maintain a mechanism for individuals to conveniently opt-out from the AI functionality in favor of a human alternative, where practicable and consistent with applicable law and governmentwide guidance.

The discussion at our workshop strongly indicated that minimum practices such as these are unlikely to be achieved unless they are made part of the procurement process and built into the system design from the outset. The OMB missed an opportunity when it failed to say so explicitly in its memo.

The memo goes on to directly address procurement. Its recommendations to agencies for responsible procurement of AI include:

**Transparency and Performance Improvement**. Agencies should take steps to ensure transparency and adequate performance for their procured AI, including by:

> A. obtaining adequate documentation to assess the AI's capabilities, such as through the use of model, data, and system cards;
> B. obtaining adequate documentation of known limitations of the AI and any guidelines on how the system is intended to be used;
> C. obtaining adequate information about the provenance of the data used to train, fine-tune, or operate the AI;
> D. regularly evaluating claims made by Federal contractors concerning both the effectiveness of their AI offerings as well as the risk management measures put in place, including by testing the AI in the particular environment where the agency expects to deploy the capability;
> E. considering contracting provisions that incentivize the continuous improvement of procured AI; and
> F. requiring sufficient post-award monitoring of the AI, where appropriate in the context of the product or service acquired.

All of these disclosures, unless they are rendered inaccessible by intellectual property claims of vendors, could be useful in contesting the accuracy or reliability of rights-affecting decisions made by advanced systems. However, the memo does not specify that the material must be publicly available. This is a potentially fatal flaw, if vendors were to provide the information to the government but insist that it could not be available to the individuals actually affected by the systems. Discussion at our workshop and research on contesting automated decision-making systems revealed a long pattern of vendors invoking IP to shield their systems from scrutiny.

Moreover, the list of items that government agencies should obtain does not include the actual algorithm or model that a system uses. Advocates challenging automated systems as unreliable have argued that access to the underlying algorithm or model can be crucial.

OMB will have a second chance to address contestability issues and the need for contestability by design in procurement processes. The March memo states that, consistent with section 7224(d) of the Advancing American AI Act and Section 10.1(d)(ii) of Executive Order 14110, OMB will also develop an "initial means" to ensure that federal contracts for the acquisition of an AI system or service align with the guidance provided in the AI M-memo and advance the other aims identified in the Advancing American AI Act.

In March 2024, at the same time that OMB issued its memo, the agency issued a request for information on the responsible procurement of artificial intelligence, to inform its efforts to develop the "initial means."[134] The RFI states that the initial means will address protection of privacy, civil rights, and civil liberties; the ownership and security of data and other information created, used, processed, stored, maintained, disseminated, disclosed, or disposed of by a contractor or subcontractor on behalf of the Federal Government; considerations for securing the training data, algorithms, and other components of any AI system against misuse, unauthorized alteration, degradation, or rendering inoperable; and any other issue or concern determined to be relevant by the OMB Director to ensure appropriate use and protection of privacy and government data and other information. It asks for input on the following questions:

- What access to documentation, data, code, models, software, and other technical components might vendors provide to agencies to demonstrate compliance with the requirements established in the March 2024 memo? What contract language would best effectuate this access, and is this best envisioned as a standard clause, or requirements-specific elements in a statement of work?
- Which elements of testing, evaluation, and impact assessments are best conducted by the vendor, and which responsibilities should remain with the agencies?
- What if any terms should agencies include in contracts to protect the Federal Government's rights and access to its data, while maintaining protection of a vendor's intellectual property?
- What if any terms, including terms governing information-sharing among agencies, vendors, and the public, should be included in contracts for AI systems or services to implement the March 2024 memo's provisions regarding notice and appeal (sections 5(c)(v)(D) and (E))?
- How might agencies structure their procurements to reduce the risk that an AI system or service they acquire may produce harmful or illegal content, such as fraudulent or deceptive content, or content that includes child sex abuse material or non-consensual intimate imagery?
- How might OMB ensure that agencies procure AI systems or services in a way that advances equitable outcomes and mitigates risks to privacy, civil rights, and civil liberties?

---

[134] Office of Management and Budget, Request for Information: Responsible Procurement of Artificial Intelligence in Government, 89 FR 22196 (Mar 29 2024), https://www.federalregister.gov/documents/2024/03/29/2024-06547/request-for-information-responsible-procurement-of-artificial-intelligence-in-government.

The answer to these questions could determine the availability of a meaningful right to contest for years to come.

The General Services Administration offers an AI governance toolkit[135] it describes as "a living and evolving guide to the application of Artificial Intelligence for the U.S. federal government." It is intended to help leaders understand what to consider as they invest in AI and lay the foundation for its enterprise-wide use. It helps leaders understand the types of problems that are best suited for the application of AI technologies, think through the building blocks they require to take advantage of AI, and how to apply AI to use cases at the project level. The entire guide references contestability and appeal rights only once, as a question to be asked ("Does the AI system provide clear notice of its use to impacted people, including what relevant factors are important to any decisions or determinations? Is there a mechanism for impacted people to contest, correct, or appeal or even opt out of the use of an AI system?") without any indication of how to ensure contestability.

At the state level as well, procurement policies so far have not made contestability a design feature for advanced automated systems. In March 2024, the state of California issued two documents specifically focused on state government use of generative AI: guidelines for public sector procurement, uses and training[136] and a risk assessment methodology issued as a supplement to the State Information Management Manual.[137] The guidelines do not specifically identify contestability as an issue to be considered in procurements. The guidelines require state entity directors and their executive leadership teams, including their CIOs, to undergo

---

[135] Artificial Intelligence Center of Excellence, General Services Administration, *AI Guide for Government*, https://coe.gsa.gov/coe/ai-guide-for-government/print-all/index.html.

[136] *State of California GenAI Guidelines for Public Sector Procurement, Uses and Training*

(March 2024) https://www.govops.ca.gov/wp-content/uploads/sites/11/2024/03/3.a-GenAI-Guidelines.pdf. The guidelines define generative AI as:

> Pretrained AI models that can generate images, videos, audio, text, and derived synthetic content. GenAI does this by analyzing the structure and characteristics of the input data to generate new, synthetic content similar to the original. Decision support, machine learning, natural language processing/translation services, computer vision and chatbot technologies or activities support may be related to GenAI, but they are not GenAI on their own.

[137] State of California Department of Technology Office of Information Security, *Generative Artificial Intelligence Risk Assessment,* SIMM 5305-F (March 2024) https://cdt.ca.gov/wp-content/uploads/2024/03/SIMM-5305-F-Generative-Artificial-Intelligence-Risk-Assessment-FINAL.pdf.

GenAI training and to review annual employee training and policy to ensure staff understand and acknowledge the acceptable use of GenAI tools. The guidelines go on to require that senior officials identify a business need and understand the implications of using GenAI to solve that problem statement before procuring new GenAI technology; assess the risks and potential impacts of deploying the GenAI under consideration; prepare data inputs and test models adequately; and establish a GenAI-focused team responsible for continuously evaluating the potential use of GenAI and its implications for operations and program administration. The guidelines require consultation with state employee end users but not with affected residents. The guidelines were issued on an interim basis as the state seeks to publish a final procurement and training policy in 2025, after extensive piloting, research and stakeholder engagement.

The risk assessment addition to the information management manual is more detailed, but it too does not require that systems be designed to ensure contestability. It starts from the proposition that public services should not be solely contingent upon GenAI systems. The document lays out a risk assessment process required for all new GenAI procurements and acquisitions, then states, "However, it is up to the entity to identify their risk tolerance and apply risk mitigation strategies that align with their organizational acceptable risk standards."

Under the manual, each GenAI use case or system must be assigned a risk level: high, moderate, or low. For all GenAI systems where the risk level is rated moderate or high, the manual requires certain mandatory quality and security and safety controls. All controls must be met. They include, "The GenAI system will have human verification to ensure accuracy and factuality of the output." But human in the loop is different from an opportunity to challenge the system's output. Moreover, if a system is operating at any scale, human verification of the accuracy and factuality of every output is simply impossible. The assessment asks whether the output of the system will make decisions that impact access to, or approval for, housing or accommodations, education, employment, credit, health care, or criminal justice, what mechanism will the GenAI system use to notify a user that they are interacting with a GenAI system rather than a human, and how will customers receive an output, and what is the mechanism to correct or appeal an error. It does not specify minimum standards for transparency or appeal. Indeed, the document allows as one option: "System operates automatically with no human intervention." Regarding intellectual property, it requires that all generated output will be owned by the State of California, but it does not address access to the underlying models, even if those models were improved using state data.

# Chapter 8: In Conclusion

The workshop's conclusion was a discussion of "What principles should guide the development of advanced automated decision making for use by state and federal governments?" The rich conversation that ensued was instrumental in developing the recommendations that form Part I of this report.[138]

During the two-day meeting, we heard of multiple examples of where complex user needs were not addressed by the advanced automated systems already in use. One of the crucial points articulated during the workshop was the extent to which advanced automated systems involved in decision making about individuals are sociotechnical systems. It follows that integrating the human, social, organizational. and technical aspects into the system from the initial design planning stages on is crucial.[139]

Discussion during the meeting made clear that this requires the right people be included in all phases of the development an advanced automated decision-making system: design, development, procurement, testing, and use. The "right people" include social scientists, lawyers and policy experts, technologists, and stakeholders who will be directly affected by the system (both government agency members and individuals using the system), the latter including a wide variety of users from various demographics and backgrounds impacted by the systems.  Because of its impact on people, the system must be stress tested in real-world cases prior to deployment. These points are reflected in Recommendations 4-7.

One particular outcome follows: It can be the case that certain advanced automated technologies are simply not appropriate to be used in systems makings decisions about individuals. For even though the requirement for contestability in government decision-making systems about individuals is not absolute,[140] the obligation is to be taken very seriously. It is thus sufficiently important to potentially rule out some technical solutions in certain uses. This underscores the need to make such a determination early in a design process for an automated system as well as the need to have all relevant stakeholders in the room from the initial planning stages of an advanced automated decision-making system.

---

[138] We remind the reader that although the workshop discussions informed the recommendations that appear in Part 1, the workshop organizing committee—Steven M. Bellovin, James X. Dempsey, Ece Kamar, and Susan Landau—are the authors and bear sole responsibility of that work.

[139] See, e.g., Gordon Baxter and Ian Sommerville, *Socio-technical systems: From design methods to systems engineering,* 23 Interacting with computers, 4 (2011).

[140] See Chapter 2 for more details.

# Acknowledgements

The workshop on Advanced Automated Systems, Contestability, and the Law was supported by National Science Foundation grants CNS 2349803 and 2349804 and the William and Flora Hewlett Foundation grant 2021-2970.

The organizers of the report are deeply grateful to the following:

Robert F. Sproull, University of Massachusetts, and Jon Eisenberg, National Academies of Science, Engineering, and Medicine, for providing valuable advice on organizing such a meeting and report;

Reva Schwartz of the National Institute of Standards and Technology for invaluable help in suggesting topics, directions, and people to include to the meeting;

Robert Pool, for distilling the summary from two days of complex discussions;

Deirdre K. Mulligan, Jeremy Epstein, and Alan Mislove of the Office of Science and Technology Policy for providing useful suggestions as we organized the meeting;

Sara Kiesler of the National Science Foundation for aiding and guiding us from an August 2023 conception of the meeting to an actual funded workshop in January 2024;

Tracey Ziegler of the National Science Foundation for ensuring that the meeting, held at the agency's Alexandria's facilities, ran smoothly;

Joshua Anderson of Tufts University, who handled all the logistics with his usual skill, grace, and aplomb, despite the complexities of holding a meeting in a facility that none of the meeting organizers had seen and during a time when Covid continues to cause changes in best-laid plans;

Shaunna Francis of Tufts University for doing a great job of typesetting;

The participants, who provided the ideas, insights, and knowledge that enabled the writing of this report.[141]

---

[141] Responsibility for the recommendations and the content of this report rests solely, however, with the organizing committee: Steven M. Bellovin, James X. Dempsey, Ece Kamar, and Susan Landau.

# Appendix:
# List of Participants[142]

**Michael Akinwumi**
National Fair Housing Alliance

**Vikram Barad**
Internal Revenue Service

**Steven Bellovin**,
 Columbia University

**Michael Berkholtz**
General Services Administration

**Miranda Bogen**
Center for Democracy & Technology

**Elizabeth Bond**
Consumer Financial Protection Bureau

**Sylvia Butterfield**
National Science Foundation

**Joseph Calandrino**
Federal Trade Commission

**Dilma Da Silva**
National Science Foundation

**Joanna Darcus**
Department of Justice

**Hal Daumé III**
University of Maryland

**Kevin De Liban**
TechTonic Justice

**Benjamin DellaRocca**
Office of Science and Technology Policy

**James X. Dempsey**
University of California Berkeley/Stanford University

**Richard Eppink**
Wrest Legal Collective

**Jeremy Epstein**
Office of Science and Technology Policy

**Emma Fedison**
Meta

**Alexandra Givens**
Center for Democracy & Technology

**Janet Haven**
Data & Society

**Michael Hawes**
Census Bureau

---

[142] The workshop participants provided ideas, insights, and knowledge to discussions on contestability in government advanced automated systems. The recommendations are due to the organizing committee: Steven M. Bellovin, James X. Dempsey, Ece Kamar, and Susan Landau.

**Kristyn Herbert**
Idaho disability rights advocate

**Susan Hirsch**
National Science Foundation

**Eunice Ikene**
Equal Employment Opportunity Commission

**Arnav Jagasia**
Palantir

**Ece Kamar**
Microsoft

**Luke Keller**
Census Bureau

**Sara Kiesler**
National Science Foundation

**Hima Lakkaraju**
Harvard University

**Susan Landau**
Tufts University

**Zachary Lipton**
Carnegie Mellon University

**Gary Lopez**
Microsoft

**Yeshimabeit Milner**
Data 4 Black Lives

**Alan Mislove**
Office of Science and Technology Policy

**Jumana Musa**
National Association of Criminal Defense Lawyers

**Reza Rashidi**
Internal Revenue Service

**Gabriel Ravel**
California Government Operations Agency

**Catherine Sharkey**
New York University

**Paul Shute**
U.S. Veterans Affairs Department

**Sid Sinha**
Social Security Administration

**Richard Speidel**
General Services Administration

**Diane Staheli**
Office of Science and Technology Policy

**Martin Stanley**
National Institute of Standards and Technology

**Sam Tyner-Monroe**
DLA Piper

**Kush Varshney**
IBM

**Lauren Wilcox**
eBay

**Valerie Szczepanik**
U.S. Securities and Exchange Commission